\documentclass[a4paper,12pt]{article}

\usepackage[left=25mm, right=25mm, top=30mm, bottom=30mm]{geometry}
\usepackage[utf8]{inputenc}
\usepackage[english]{babel}
\usepackage{amsmath}
\usepackage{amsfonts}
\usepackage{mathrsfs}
\usepackage{dsfont}
\usepackage{simplewick}
\usepackage{graphicx}
\usepackage{cite}
\usepackage[colorlinks, allcolors=blue, linktocpage=true]{hyperref}

\renewcommand{\O}{\mathcal{O}}
\newcommand{\ket}[1]{\left| #1 \right\rangle}
\newcommand{\bra}[1]{\left\langle #1 \right|}

\newcommand{\SO}{\text{SO}}

\newcommand{\Deltatilde}{\widetilde{\Delta}}
\newcommand{\lambdatilde}{\widetilde{\lambda}}
\DeclareMathOperator{\T}{T}
\DeclareMathOperator{\Tbar}{\overline{T}}
\DeclareMathOperator{\re}{Re}
\DeclareMathOperator{\im}{Im}

\makeatletter
\newsavebox{\@brx}
\newcommand{\llangle}[1][]{\savebox{\@brx}{\(\m@th{#1\langle}\)}%
  \mathopen{\copy\@brx\kern-0.5\wd\@brx\usebox{\@brx}}}
\newcommand{\rrangle}[1][]{\savebox{\@brx}{\(\m@th{#1\rangle}\)}%
  \mathclose{\copy\@brx\kern-0.5\wd\@brx\usebox{\@brx}}}
\makeatother

\title{Conformal 3-point functions and \\
the Lorentzian OPE in momentum space}

\author{Marc Gillioz}

\date{%
SISSA, via Bonomea 265, 34136 Trieste, Italy \\[3mm]
and \\[3mm]
Theoretical Particle Physics Laboratory, Institute of Physics, \\
EPFL, Lausanne, Switzerland
}

\begin{document} 
\maketitle

\vspace{1cm}

\begin{abstract}
In conformal field theory in Minkowski momentum space, the 3-point correlation functions of local operators are completely fixed by symmetry. Using Ward identities together with the existence of a Lorentzian operator product expansion (OPE), we show that the Wightman function of three scalar operators is a double hypergeometric series of the Appell $F_4$ type. We extend this simple closed-form expression to the case of two scalar operators and one traceless symmetric tensor with arbitrary spin. Time-ordered and partially-time-ordered products are constructed in a similar fashion and their relation with the Wightman function is discussed.
\end{abstract}

\newpage
\tableofcontents


\section{Introduction}

Conformal field theory can be formulated algebraically in terms of a set of primary operators and of rules that define the operator product expansion (OPE).
Equivalently, all correlation functions of a conformal field theory can be obtained from 2- and 3-point functions, which are themselves fixed by conformal symmetry up to a small number of numerical coefficients.
This statement applies both to correlation functions in position space and in momentum space, but it is by far more common in conformal field theory to use the position-space representation. There are several good reasons why it is so:
\begin{enumerate}

\item[(1)]
All the 2- and 3-point functions in position space are known and relatively easy to evaluate.
In the case of scalar operators and of operators carrying low-dimensional spin representations they have been known since a long time~\cite{Polyakov:1970xd, Schreier:1971um, Osborn:1993cr, Erdmenger:1996yc}. More recently, correlation functions involving larger spin representations have been constructed using the embedding-space formalism \cite{Giombi:2011rz, Costa:2011mg, Maldacena:2011jn, Zhiboedov:2012bm, Elkhidir:2014woa, Fortin:2019xyr, Fortin:2019pep}, and this construction can be extended in an algorithmic way to arbitrary spin representations using weight-shifting operators \cite{Karateev:2017jgd, Karateev:2018oml}.

\item[(2)]
Higher-point functions can be computed with the help of an operator product expansion that has a large range of convergence.
This OPE applies naturally when two operators are close in space, but it actually extends over most of the possible configuration space~\cite{Pappadopulo:2012jk, Hogervorst:2013sma, Rychkov:2015lca, Mukhametzhanov:2018zja}.
This implies in particular that distinct convergent OPEs can be used to compute the same correlation function, which is the key property exploited by the conformal bootstrap~\cite{Rattazzi:2008pe, Rychkov:2016iqz, Simmons-Duffin:2016gjk, Poland:2018epd}.

\item[(3)]
There is a simple connection between the correlation function in Euclidean and Minkowski position space: Wightman functions in Minkowski space are obtained from Euclidean correlators by a straightforward Wick rotation. This property relates unitary Lorentzian theories to reflection-positive Euclidean ones and explains the reality of 3-point function coefficients.

\end{enumerate}

\noindent
All of these properties are altered in momentum space. Let us review them in reverse order:
\begin{enumerate}

\item[(3)]
There are branch cuts in the complexified momentum space that make the Wick rotation between Minkowski and Euclidean space non-trivial. A comprehensive discussion of this phenomenon has recently appeared in ref.~\cite{Bautista:2019qxj}.
We shall see in this paper that the time-ordered product of operators in Minkowski momentum space is simply related to the known Euclidean expression, but also that the Wightman functions are very different objects that do not have a Euclidean counterpart.

\item[(2)]
A momentum-space OPE can be defined by the Fourier transform of the position-space OPE.
Its convergent limit is when the two operators involved have both large momenta~\cite{Coriano:2013jba}. 
However, this momentum-space OPE has only been formulated in Euclidean theories so far. Very little is known about the Lorentzian OPE, about its convergence properties, or even whether it converges at all.

\item[(1)]
Maybe more surprisingly, our knowledge of conformal 3-point functions in momentum space is quite incomplete.
While they have been extensively studied in Euclidean theories~\cite{Coriano:2013jba, Bzowski:2013sza, Bzowski:2015pba, Coriano:2017mux, Bzowski:2017poo, Isono:2018rrb, Bzowski:2018fql, Coriano:2018bsy, Isono:2019ihz, Isono:2019wex},
partly because of their relevance for inflation~\cite{Maldacena:2011nz, Creminelli:2012ed, Kundu:2014gxa, Arkani-Hamed:2015bza, Kundu:2015xta, Shukla:2016bnu, Arkani-Hamed:2018kmz, Sleight:2019mgd, Sleight:2019hfp},
it is only recently that their study in Lorentzian signature has begun~\cite{Gillioz:2016jnn, Gillioz:2018kwh, Gillioz:2018mto, Bautista:2019qxj}.
Moreover, even in the simplest case of 3 scalar operators, the only expression available in the literature so far is in the form of a quite complicated integral over Bessel functions~\cite{Bautista:2019qxj}.

\end{enumerate}

\noindent
In spite of these difficulties, there exist strong motivations to study conformal field theory in Minkowski momentum space. For instance, the light-like limit of momentum-space correlators is intrinsically connected with the study of local operators integrated along a light ray~\cite{Belitsky:2013xxa, Kravchuk:2018htv, Kologlu:2019bco, Kologlu:2019mfz}, which has been instrumental in the derivation of conformal collider bounds~\cite{Hofman:2008ar, Cordova:2017zej, Cordova:2017dhq, Chowdhury:2017vel, Afkhami-Jeddi:2017rmx, Afkhami-Jeddi:2018own, Meltzer:2018tnm, Belin:2019mnx}, of the proof of the average null energy condition from causality~\cite{Hofman:2009ug, Hartman:2016lgu}
or even in the study of asymptotic symmetries~\cite{Cordova:2018ygx}.
A limit of the momentum space 3-point function also enters in the light-cone Hamiltonian truncation formalism~\cite{Katz:2016hxp, Fitzpatrick:2018ttk},
and the use of momentum space makes anomaly coefficients appear explicitly in correlators~\cite{Gillioz:2016jnn, Gillioz:2018kwh}.
Moreover, when the momentum-space 3-point functions are known, constructing conformal blocks out of them is simple in the sense that it does not require additional integration~\cite{Gillioz:2016jnn, Gillioz:2018kwh, Gillioz:2018mto}:
a recent example where this technology has been put to good use is ref.~\cite{Erramilli:2019njx}.

Even though this list of motivations is far from exhaustive, it makes evident that there is an interest in closing the gaps of points (1) and (2) discussed above. The goal of this paper is precisely to improve on point (1) by providing a simple closed-form expression for the Wightman 3-point function in Minkowski momentum space.
This goal is after all quite modest since it consists in taking the Fourier transform of a known position-space 3-point function, but we will see that its computation is not quite simple. 
Along the way, we will also touch upon point (2), although without discussing the delicate issue of OPE convergence.

\subsection{Strategy and main result}

The strategy for determining the 3-point function will be to use conformal Ward identities to express it in terms of the solutions of some differential equation, in the spirit of the Euclidean derivation of ref.~\cite{Coriano:2013jba}.
What is new in our case is not so much the difference between the Euclidean and Minkowskian conformal algebras as it is the boundary condition provided by the Lorentzian OPE.
With this strategy, we do not perform directly the Fourier transform of the position-space correlation function, although we use it to verify numerically and in some limits analytically the validity of our derivation.
In the case of 3 scalar operators, our result is
\begin{equation}
	\llangle \phi_f(p_f) \phi_0(p_0) \phi_i(p_i) \rrangle
	= \lambdatilde_{f0i}
	\Theta(-p_f) \Theta(p_i)
	\frac{(-p_f^2)^{\Delta_f - d/2} (-p_i^2)^{\Delta_i - d/2}}{(p_0^2)^{(\Delta_i + \Delta_f - \Delta_0)/2}}
	F_{\Delta_f \Delta_0 \Delta_i}\left( \frac{p_f^2}{p_0^2}, \frac{p_i^2}{p_0^2} \right),
\label{eq:mainresult}
\end{equation}
where $F_{\Delta_f \Delta_0 \Delta_i}$ is an Appell $F_4$ generalized hypergeometric function of two variables defined in eq.~\eqref{eq:F},
and $\lambdatilde_{f0i}$ is an OPE coefficient related to the usual one by eq.~\eqref{eq:lambdatilde}.
We have eliminated the $\delta$-function demanding overall momentum conservation by use of the notation
\begin{equation}
	\bra{0} \O_1(p_1) \cdots \O_n(p_n) \ket{0}
	\equiv (2\pi)^d \delta^d(p_1 + \ldots + p_n)
	\llangle \O_1(p_1) \cdots \O_n(p_n) \rrangle,
\label{eq:notation:deltafunction}
\end{equation}
and the $\Theta$-functions impose conditions on the momenta $p_i$ and $p_f$,%
\footnote{Note that this definition is equivalent to $\Theta(p) = \Theta(p^0 - |\vec{p}|)$ used in ref.~\cite{Bautista:2019qxj}, but we write it as a function of $p^2$ and $p^0$ to emphasize the fact that it is a Lorentz-invariant object.}
\begin{equation}
	\Theta(p) \equiv \Theta(-p^2) \Theta(p^0)
	= \left\{ \begin{array}{ll}
		1 \quad & \text{if}~p~\text{is time-like with positive energy},
		\\
		0 & \text{otherwise.}
	\end{array} \right.
\label{eq:Theta}
\end{equation}
This result applies in the regime $|p_f|, |p_i| < |p_0|$ where the function $F_{\Delta_f \Delta_0 \Delta_i}$ is analytic,
while the discussion of section~\ref{subsec:analyticcontinuation} and in particular the general expression~\eqref{eq:3scalars:complete} covers all kinematic configurations.
This result is valid for any scaling dimensions of the operators, as well as in any space-time dimension $d \geq 2$.

The rest of the paper is organized as follows: Section~\ref{sec:scalar} is devoted to explaining the steps that lead to the result~\eqref{eq:mainresult}. It also contains discussions of the special cases that are generalized free field theory and $d = 2$ space-time dimensions.
We then generalize this result in section~\ref{sec:spinning}, replacing one of the scalar operators with a traceless symmetric tensor of arbitrary spin.
In section~\ref{sec:timeordering} we introduce the time-ordering operator in the 3-point function, show how the result differ from the Wightman function, and perform consistency checks.
The appendix presents the direct Fourier transform of the position-space correlation function used to verify our results numerically.


\section{The Wightman 3-point function of scalar operators}
\label{sec:scalar}

We begin with a derivation of the Wightman function of 3 distinct scalar operators. The momentum-space representation of this correlation function could in principle be obtained directly by performing the Fourier transform of the Wightman function in position space, which is known and relatively simple. This is the approach followed in ref.~\cite{Bautista:2019qxj}, and the result is an integral over Bessel functions. We will follow instead a different approach purely based on the symmetries of the 3-point function and on the existence of an operator product expansion (OPE).
Our result is a closed-form expression, which provides a more practical and efficient way of evaluating the scalar 3-point function at any point in momentum space.

\subsection{Momentum eigenstates and support}
\label{subsec:momentumeigenstates}

Before we begin with the derivation, it is useful to recall some properties of the momentum-space representation.
The Hilbert space of a conformal field theory can be constructed in terms of a (infinite) set of primary states $\ket{\O}$ and of their descendants obtained by acting repeatedly with the generator of translations $P^\mu$,
\begin{equation}
	\ket{ \O },
	\quad
	P^\mu \ket{ \O },
	\quad
	P^2 \ket{ \O },
	\quad
	\left( P^\mu P^\nu - \tfrac{1}{d} \eta^{\mu\nu} P^2 \right) \ket{ \O },
	\quad
	\ldots
\end{equation}
An equivalent representation of this Hilbert space is in terms of distributions over flat Minkowski space,
\begin{equation}
	\ket{ \O(x) } \equiv e^{-i \, x \cdot P} \ket{ \O },
	\qquad
	x \in \mathbb{R}^{d-1,1}.
\label{eq:states:position}
\end{equation}
The state/operator correspondence associates to each such state a local operator $\O(x)$ such that
\begin{equation}
	\O(x) \ket{0} = \ket{ \O(x) }.
\end{equation}
Taking $P^0$ as the Hamiltonian of the theory, one should include in the definition \eqref{eq:states:position} the prescription $x^0 \to x^0 + i \epsilon$ with positive $\epsilon$ so that the norm of the state $\ket{ \O(x) }$ is well-defined when the Hamiltonian is bounded from below.%
\footnote{The Minkowski metric is taken in the ``mostly $+$'' convention, i.e.~$\eta = \text{diag}(-1, 1, \ldots, 1)$
and Lorentz indices run from $0$ to $d - 1$.}
For a scalar state $\ket{ \phi(x)}$, this norm, or equivalently the Wightman 2-point function of the operator $\phi(x)$, is given by
\begin{equation}
	\langle \phi(x_1) | \phi(x_2) \rangle
	= \bra{0} \phi(x_1) \phi(x_2) \ket{0}
	= \frac{1}{[ -(x_1^0 - x_2^0 - i \epsilon)^2 + (\vec{x}_1 - \vec{x}_2)^2  ]^\Delta},
\label{eq:norm:position}
\end{equation}
where $\Delta$ is the scaling dimension of the operator $\phi$ and the normalization is conventionally chosen.

There exist yet another equivalent representation of the Hilbert space given by the states
\begin{equation}
	\ket{ \O(p) } \equiv \int d^dx \, e^{i \, p \cdot x} \ket{ \O(x) },
\label{eq:state:momentum}
\end{equation}
which are eigenstates of the generator of translations, $P^\mu \ket{ \O(p) } = p^\mu \ket{ \O(p) }$.
As before, these states are in one-to-one correspondence with the set of operators
\begin{equation}
	\O(p) \equiv \int d^dx \, e^{i \, p \cdot x} \O(x).
\label{eq:operator:momentum}
\end{equation}
The advantage of this basis is that the states are orthogonal in a distributional sense: their norm satisfies%
\footnote{Our definition of the conjugate state is $\bra{ \O(p) } = \bra{0} \O(p)$ and therefore $\ket{ \O(p) }^\dag = \bra{ \O(-p) }$.}
\begin{equation}
	\langle \phi(p_f) | \phi(p_i) \rangle
	= (2\pi)^d \delta^d(p_f + p_i)
	\Theta(p_i)
	\frac{2^{d- 2 \Delta + 1} \pi^{(d+2)/2}}
	{\Gamma\left( \Delta \right) \Gamma\left( \Delta - \frac{d-2}{2} \right)}
	(-p_i^2)^{\Delta - d/2},
\label{eq:norm:momentum}
\end{equation}
and hence vanishes if $p_f \neq - p_i$. The function $\Theta$ is defined in eq.~\eqref{eq:Theta}: it indicates that the norm only has support when $p_i$ (and thus $-p_f$) is time-like and has positive energy.
When this condition is not satisfied, the state must be null:
\begin{equation}
	\ket{ \phi(p) } = 0
	\qquad
	\text{if}~ p^2 > 0 ~\text{or}~ p^0 < 0.
\label{eq:nullstates}
\end{equation}
Note that this property is specific to the Lorentzian theory: momentum eigenstates can be constructed in an Euclidean theory but they have different characteristics.
In the notation of eq.~\eqref{eq:notation:deltafunction}, the Wightman 2-point function of a scalar operator is therefore
\begin{equation}
	\llangle \phi(-p) \phi(p) \rrangle
	= \Theta(p)
	\frac{2^{d- 2 \Delta + 1} \pi^{(d+2)/2}}
	{\Gamma\left( \Delta \right) \Gamma\left( \Delta - \frac{d-2}{2} \right)}
	(-p^2)^{\Delta - d/2}
	\equiv W_\Delta(p).
\label{eq:Wightman2ptfct}
\end{equation}

\begin{figure}
	\centering
	\footnotesize
	\begin{tabular}{c@{\hspace{1.5cm}}c}
		\includegraphics[width=0.4\linewidth]{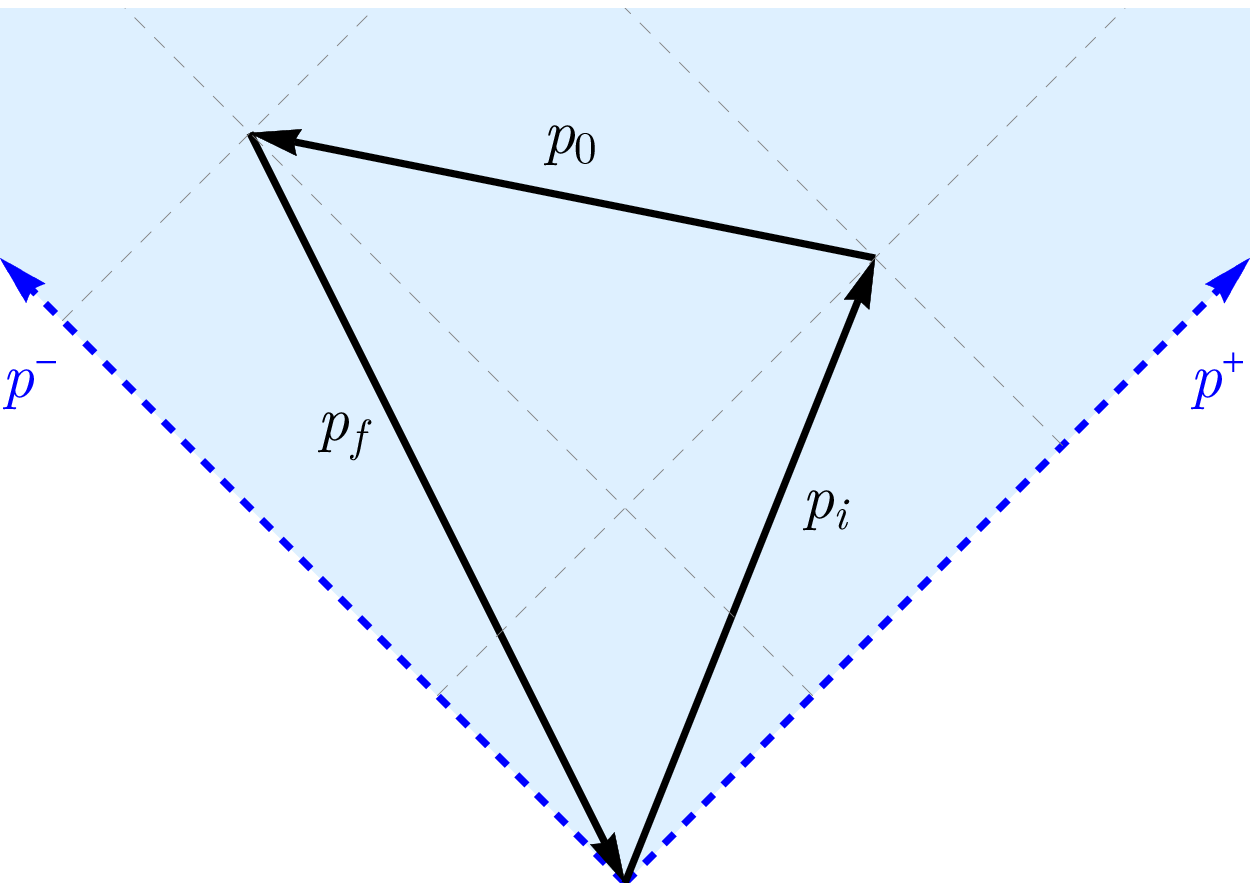}
		&
		\includegraphics[width=0.4\linewidth]{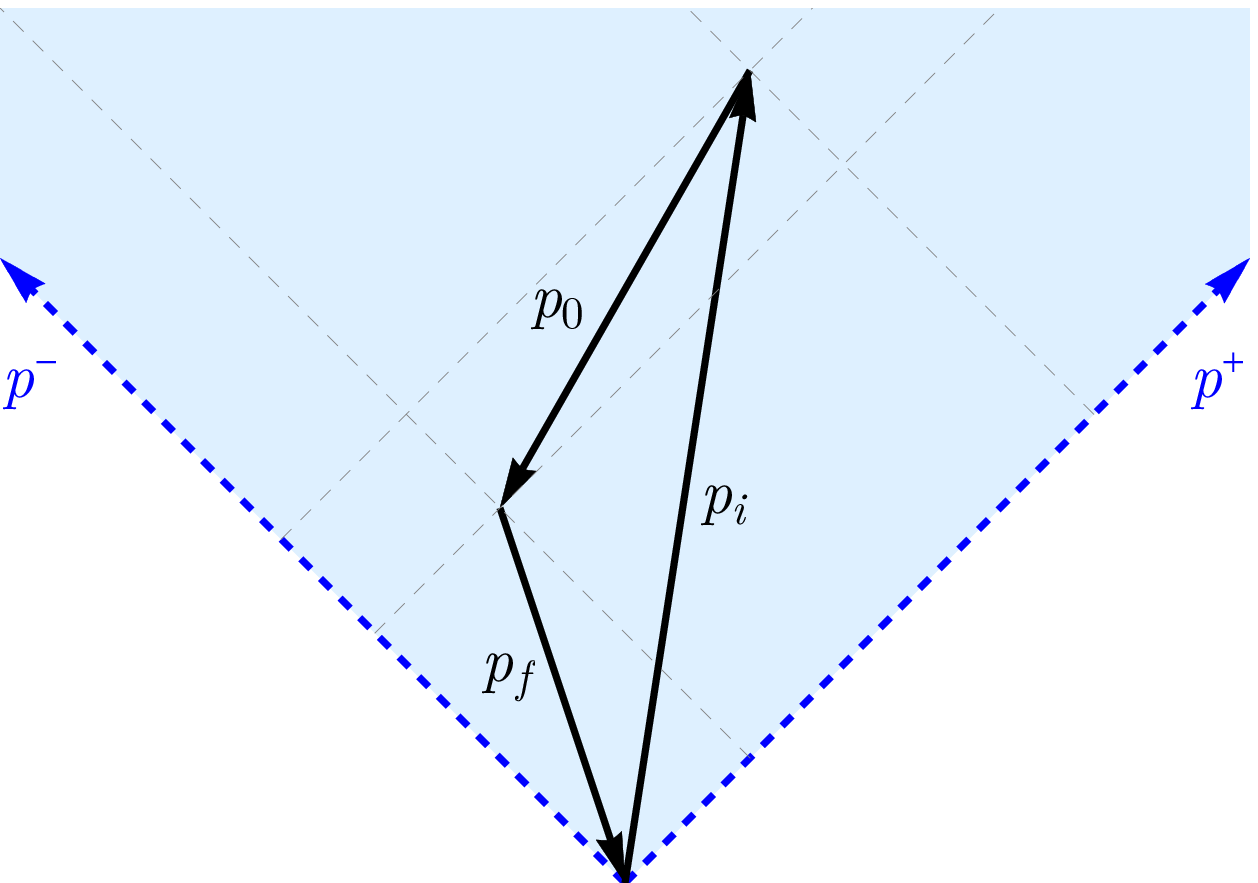}
		\\[1em]
		(a) $p_0^2 > 0$ & (b) $p_0^2 < 0$
	\end{tabular}
	\caption{Two examples of momentum configurations for the Wightman 3-point function.
	The momenta $p_i$, $p_0$ and $p_f$ add up to zero, and both $p_i$ and $-p_f$ must lie in the light cone indicated in blue for the 3-point function to be non-zero. The intermediate momentum $p_0$ can either be space-like (a) or time-like (b).}
	\label{fig:kinematics}
\end{figure}

These general considerations are also important for the Wightman 3-point function since it can be written as the expectation value of an operator between two momentum eigenstates,
\begin{equation}
	\bra{0} \phi_f(p_f) \phi_0(p_0) \phi_i(p_i) \ket{0}
	= \bra{ \phi_f(p_f) } \phi_0(p_0) \ket{ \phi_i(p_i) },
\end{equation}
where we have used the labels $i$ for ``initial'' and $f$ for ``final'' states. Because of the condition \eqref{eq:nullstates} on the states, this 3-point function only has support when both momenta $p_i$ and $-p_f$ are time-like with positive energies, i.e.
\begin{equation}
	\llangle \phi_f(p_f) \phi_0(p_0) \phi_i(p_i) \rrangle \propto \Theta(-p_f) \Theta(p_i).
\label{eq:support}
\end{equation}
By translation invariance, correlation functions in momentum space are always proportional to a $\delta$-function, in this case enforcing $p_f + p_0 + p_i = 0$. Nevertheless, the constraint \eqref{eq:support} does not restrict the intermediate momentum $p_0$, which can be either space-like or time-like, with positive or negative energy. Two possible configurations of momenta are shown in figure~\ref{fig:kinematics}.

There are additional constraints coming from conformal symmetry: using Lorentz symmetry, we can choose to parametrize the 3-point function in terms of the three invariant quantities $p_f^2$, $p_0^2$ and $p_i^2$. This choice is not unique, but it will turn out to be the most convenient in the next sections.
The scale symmetry determines the overall scaling dimension of the 3-point function. Taking $p_0^2$ as the reference scale, we can write
\begin{equation}
	\llangle \phi_f(p_f) \phi_0(p_0) \phi_i(p_i) \rrangle
	= \Theta(-p_f) \Theta(p_i) (p_0^2)^{(\Delta_f + \Delta_0 + \Delta_i - 2d)/2}
	F\left( \frac{p_f^2}{p_0^2}, \frac{p_i^2}{p_0^2} \right).
\label{eq:scalar3ptfct}
\end{equation}
where $F$ is a function of two dimensionless arguments.
This might seem a curious choice of reference scale since both $-p_f^2$ and $-p_i^2$ are positive over the region of support whereas $p_0^2$ can potentially change sign. We will see in the next section that this choice is motivated by the various OPE limits of the 3-point function. Moreover, note that the scalar 3-point function enjoys the conjugation symmetry
\begin{equation}
	\llangle \phi_f(p_f) \phi_0(p_0) \phi_i(p_i) \rrangle
	= \llangle \phi_i(-p_i) \phi_0(-p_0) \phi_f(-p_f) \rrangle,
\label{eq:scalar3pt:conjugation}
\end{equation}
which means that the choice \eqref{eq:scalar3ptfct} makes $F$ a symmetric function under the simultaneous exchange of its two arguments and of the scaling dimensions $\Delta_f$ and $\Delta_i$.
Finally, there are constraints coming from the special conformal symmetry that will completely restrict the form of $F$. Since these constraints are much more involved, we dedicate section~\ref{subsec:conformalWardidentities} to their study.
But before proceeding with them, we discuss the role played by the operator product expansion.

\subsection{OPE limits in momentum space}
\label{subsec:OPElimits}

Besides the Hilbert space construction discussed in the previous section, the other key property of conformal field theory is the existence of an operator product expansion.
The OPE expresses how a local operator acts on the Hilbert space of the theory: in the position-space representation,
\begin{equation}
	\phi_1(x_1) \ket{ \phi_2(x_2) } = \sum_{\O} \lambda_{\O12} \,
	C_{\O12}( x_1 - x_2, P) \ket{ \O(x_2) },
\label{eq:OPE:position}
\end{equation}
where the operator $C_{\O12}(x, P)$ is completely fixed by conformal symmetry, while the OPE coefficients $\lambda_{\O12}$ encode the dynamical content of the theory.
$C_{\O12}$ is a series expansion in the generator $P^\mu$, for instance in the case where the operators $\phi_1$, $\phi_2$ and $\O$ are scalars
\begin{equation}
	C_{\O12}(x, P)
	= |x|^{\Delta_\O - \Delta_1 - \Delta_2}
	\left( 1 + \frac{\Delta_\O + \Delta_1 - \Delta_2}{2 \Delta_\O} \, x \cdot P + \ldots \right).
\label{eq:OPE:C}
\end{equation}
In an Euclidean CFT $|x|$ would be the Euclidean norm; in a Lorentzian CFT, it is its analytic continuation
$|x|^2 = -(x^0 - i\epsilon)^2 + \vec{x}^2$.

Since the OPE ultimately expresses the completeness of the Hilbert space, there must exist a similar statement in the momentum-space representation.
Taking the Fourier transform of eq.~\eqref{eq:OPE:position} with respect to both $x_1$ and $x_2$, one can write
\begin{equation}
	\phi_1(p_1) \ket{ \phi_2(p_2) } = \sum_\O \lambda_{\O12} \,
	\widetilde{C}_{\O12}(p_1, p_1 + p_2)
	\ket{ \O(p_1 + p_2) },
\label{eq:OPE:momentum}
\end{equation}
where we have defined
\begin{equation}
	\widetilde{C}_{\O12}(p, q)
	= \int d^dx \, e^{i \, p \cdot x}
	C_{\O12}(x, q).
\end{equation}
We have used the fact that $\ket{\O(p_1 + p_2)}$ is a momentum eigenstate to replace the generator $P^\mu$ by its eigenvalue. For this reason, $\widetilde{C}_{\O12}(p, q)$ is not anymore a derivative operator acting on the primary $\O$ but just a number. This is a consequence of the orthogonality of momentum eigenstates.
One should realize however that this definition of the momentum-space OPE is purely formal so far, and it faces two major problems.
First, we have not established whether the Fourier transform commutes with the sum over conformal primaries.
It is known that the OPE for Wightman functions converges in the sense of distributions~\cite{Mack:1976pa},
but the series might not converge at every given configuration of momenta (see Ref.~\cite{Gillioz:2019iye} for a discussion and examples in $d = 2$ dimensions).
In any case, this problem is absent when the OPE applies to a 3-point function since the sum is given by a single term.
The second problem is a practical one: using the expansion \eqref{eq:OPE:C}, one can formally write
\begin{equation}
	\widetilde{C}_{\O12}(p, q)
	= \left[ 1 - i \frac{\Delta_\O + \Delta_1 - \Delta_2}{2 \Delta_\O} \, 
	q^\mu \frac{\partial}{\partial p^\mu} + \ldots \right]
	\int d^dx \, e^{i \, p \cdot x} |x|^{\Delta_\O - \Delta_1 - \Delta_2},
\label{eq:OPE:Ctilde}
\end{equation}
and recognize in the integral on the right-hand side the Wightman 2-point function of a fictitious operator with scaling dimension $(\Delta_1 + \Delta_2 - \Delta_\O)/2$. This integral is discontinuous at $p^2 = 0$ and the dependence of $\widetilde{C}_{12\O}(p, q)$ on $p$ is difficult to establish, which means that this formal definition of the OPE in impractical for computations, but it establishes a property that will be crucial in the next section: by definition, $\widetilde{C}_{12\O}(p, q)$ is an analytic function in $q$ around $q = 0$.

\begin{figure}
	\centering
	\footnotesize
	\begin{tabular}{c@{\hspace{1.5cm}}c}
		\includegraphics[width=0.4\linewidth]{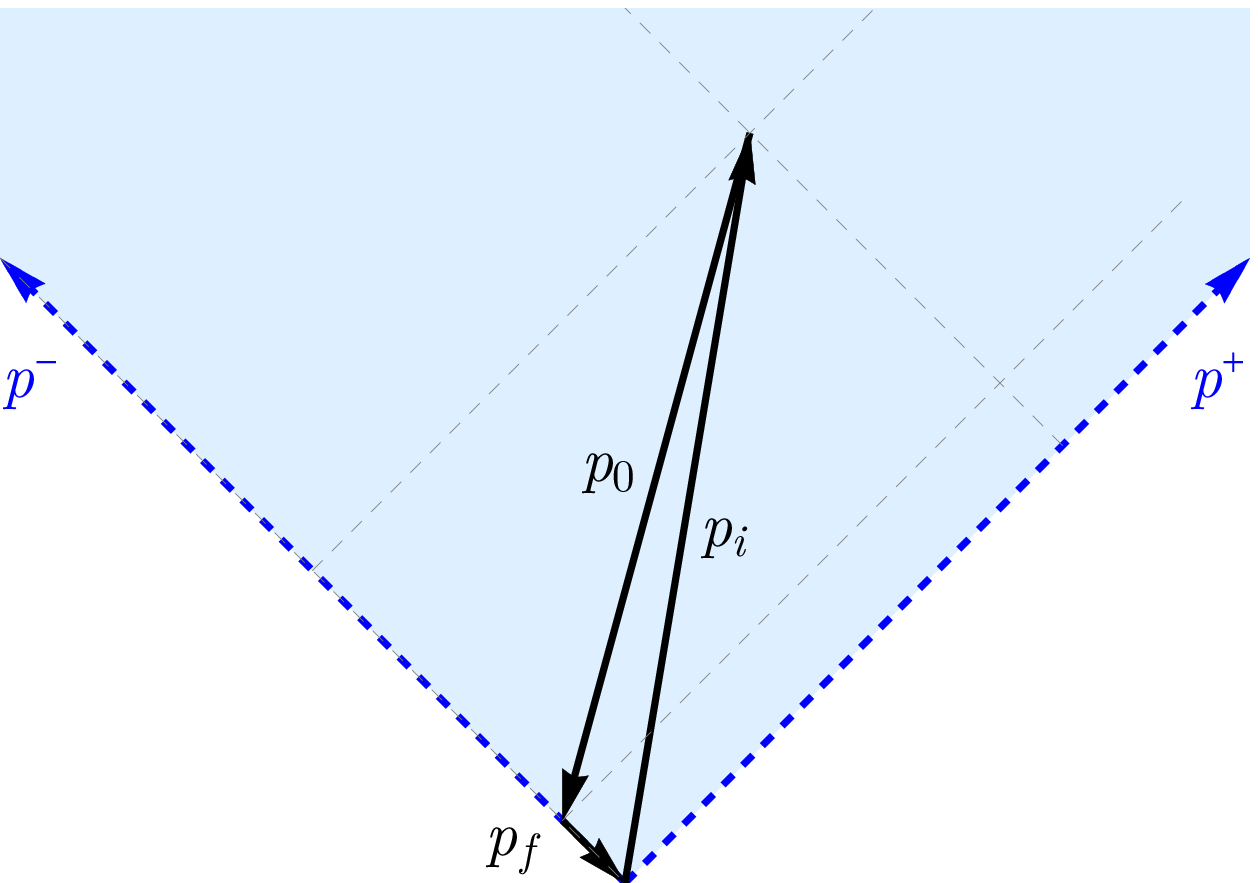}
		&
		\includegraphics[width=0.4\linewidth]{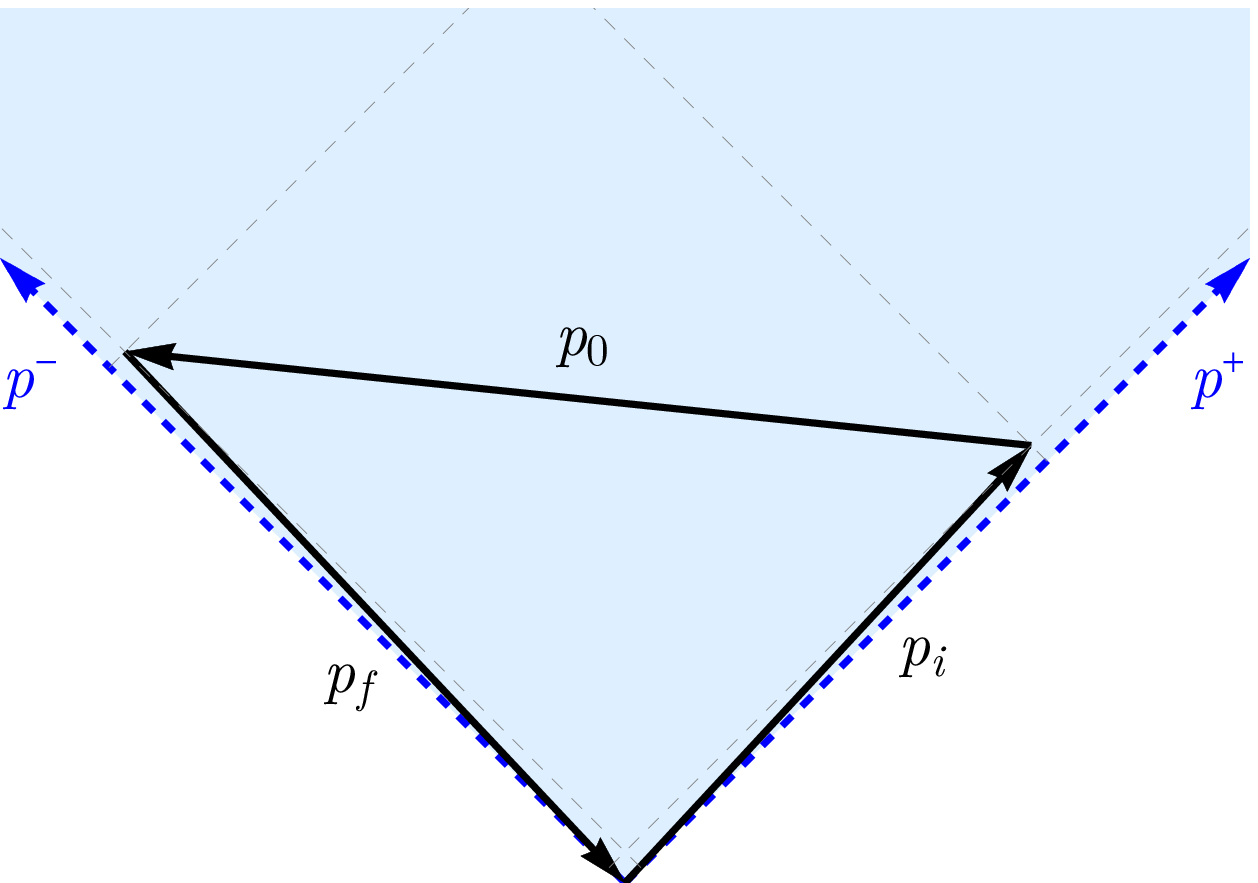}
		\\[1em]
		(a) $p_f \to 0$ & (b) $p_i^2, p_f^2 \to 0_-$
	\end{tabular}
	\caption{Examples of momentum configurations in the limits (a) $p_f \to 0$ as in eq.~\eqref{eq:limit:pf:zero} and (b) $p_f^2, p_i^2 \to 0_-$ as in eq.~\eqref{eq:limit:massless}.
	The configuration (a) also shows that the limit $p_f \to 0$ can be reached taking $p_f^2 \to 0_-$ first. 
	}
	\label{fig:kinematics:limits}
\end{figure}

Applying this momentum-space OPE to the Wightman 3-point function, one gets
\begin{equation}
	\llangle \phi_f(p_f)
	\contraction{}{\phi}{{}_0(p_0)}{\phi}
	\phi_0(p_0) \phi_i(p_i) \rrangle
	= \lambda_{f0i} \, \widetilde{C}_{f0i}( p_0, -p_f) \llangle \phi_f(p_f) \phi_f(-p_f) \rrangle
\label{eq:OPE:0i}
\end{equation}
where the line above the 3-point function indicates that the OPE is taken between $\phi_0$ and $\phi_i$.
In the limit $p_f \to 0$, the series \eqref{eq:OPE:Ctilde} for $\widetilde{C}_{f0i}( p_0, -p_f)$ is dominated by its first term, and since the integral is a Wightman 2-point function for an operator with scaling dimension $(\Delta_i + \Delta_0 - \Delta_f)/2$, we can use eq.~\eqref{eq:Wightman2ptfct} to get
\begin{equation}
	\widetilde{C}_{f0i}( p_0, 0) = \frac{2^{d- \Delta_i - \Delta_0 + \Delta_f + 1} \pi^{(d+2)/2}}
	{\Gamma\left( \frac{\Delta_i + \Delta_0 - \Delta_f}{2} \right)
	\Gamma\left( \frac{\Delta_i + \Delta_0 - \Delta_f - d + 2}{2} \right)}
	(-p_0^2)^{(\Delta_i + \Delta_0 - \Delta_f - d)/2},
\end{equation}
Note that $p_0$ is necessarily time-like in this limit since it approaches $-p_i$, as illustrated in figure~\ref{fig:kinematics:limits}~(a).
We obtain therefore the limit
\begin{equation}
\begin{aligned}
	&\llangle \phi_f(p_f) \phi_0(p_0) \phi_i(p_i) \rrangle_{p_f \to 0}
	\\
	& \qquad
	= \lambda_{f0i}
	\frac{2^{2d- \Delta_i - \Delta_0 - \Delta_f + 2} \pi^{d+2}
	(-p_f^2)^{\Delta_f - d/2}
	(-p_0^2)^{(\Delta_i + \Delta_0 - \Delta_f - d)/2}}
	{\Gamma\left( \Delta_f \right) \Gamma\left( \Delta_f - \frac{d-2}{2} \right)
	\Gamma\left( \frac{\Delta_i + \Delta_0 - \Delta_f}{2} \right)
	\Gamma\left( \frac{\Delta_i + \Delta_0 - \Delta_f - d + 2}{2} \right)}.
\end{aligned}
\label{eq:limit:pf:zero}
\end{equation}
The limit $p_i \to 0$ can established in a similar fashion starting from the OPE
\begin{equation}
	\llangle
	\contraction{}{\phi}{{}_f(p_f)}{\phi}
	\phi_f(p_f) \phi_0(p_0) \phi_i(p_i) \rrangle
	= \lambda_{f0i} \, \widetilde{C}_{i0f}( -p_0, p_i) \llangle \phi_i(-p_i) \phi_i(p_i) \rrangle.
\label{eq:OPE:f0}
\end{equation}
The result corresponds to exchanging the labels $f$ and $i$ in eq.~\eqref{eq:limit:pf:zero}.

These OPE limits are important, but in practice they will not be convenient to determine the Wightman 3-point function completely. Instead, there is another case that can be resolved with the help of the OPE: the light-cone limit $p_f^2 \to 0_-$ with $p_f \neq 0$. To understand this limit, consider that the coefficient $\widetilde{C}_{f0i}(p_0, -p_f)$ is invariant under Lorentz transformations. It can therefore be written in terms of the invariant quantities $p_f^2$, $p_0^2$ and $p_i^2$. The analyticity in $p_f$ implies that $\widetilde{C}_{f0i}(p_0, -p_f)$ is also analytic in $p_f^2$. When applied to eq.~\eqref{eq:OPE:0i}, this means that the 3-point function is equal to an analytic function of $p_f^2$ multiplying the 2-point function, and therefore
\begin{equation}
	\llangle \phi_f(p_f) \phi_0(p_0) \phi_i(p_i) \rrangle_{p_f^2 \to 0_-}
	\propto (-p_f^2)^{\Delta_f - d/2}
	(p_0^2)^{(\Delta_i + \Delta_0 - \Delta_f - d)/2}
	f\left( \frac{p_i^2}{p_0^2} \right),
	\label{eq:limit:pf2:zero}
\end{equation}
where $f$ is an unknown function. The same argument can be applied in the limit $p_i^2 \to 0$ to the OPE \eqref{eq:OPE:f0} to establish that 
\begin{equation}
	\llangle \phi_f(p_f) \phi_0(p_0) \phi_i(p_i) \rrangle_{p_i^2 \to 0_-}
	\propto (-p_i^2)^{\Delta_i - d/2}
	(p_0^2)^{(\Delta_f + \Delta_0 - \Delta_i - d)/2}
	f'\left( \frac{p_f^2}{p_0^2} \right),
\end{equation}
for a different function $f'$.
Taking both limits $p_f^2 \to 0_-$ and $p_i^2 \to 0_-$ simultaneously, in a configuration of momenta similar to figure~\ref{fig:kinematics:limits}~(b) in which $p_0$ is necessarily space-like, one must have
\begin{equation}
	\llangle \phi_f(p_f) \phi_0(p_0) \phi_i(p_i) \rrangle_{p_f^2, p_i^2 \to 0_-}
	\propto
	\frac{(-p_f^2)^{\Delta_f - d/2} (-p_i^2)^{\Delta_i - d/2}}
	{(p_0^2)^{(\Delta_f + \Delta_i - \Delta_0)/2}}.
	\label{eq:limit:massless} 
\end{equation}
This form is consistent with the ansatz~\eqref{eq:scalar3ptfct} for the Wightman 3-point function, and it establishes that the function $F$ has the asymptotic limit
\begin{equation}
	F(z_f, z_i)_{z_f, z_i \to 0_-}
	\propto (-z_f)^{\Delta_f - d/2} (-z_i)^{\Delta_i - d/2}.
	\label{eq:limit:massless:F} 
\end{equation}
When combined with the constraints from conformal Ward identities, this will completely fix the scalar Wightman function up to an overall coefficient, which in turn will be determined by the limit \eqref{eq:limit:pf:zero}.
This is the topic of the next section.

\subsection{Conformal Ward identities}
\label{subsec:conformalWardidentities}

The form \eqref{eq:scalar3ptfct} of the Wightman 3-point function already takes into account all the information from Poincar\'e and scale symmetry. Only the symmetry under special conformal transformation remains to be imposed.
To do so, we follow the approach pioneered in ref.~\cite{Coriano:2013jba}. It consists in writing down a system of differential equations for the unknown function $F$.

The infinitesimal transformations of the momentum-space operators under the conformal group are given in appendix~\ref{app:algebra}. In particular, the action \eqref{eq:primaryfieldtransformations} of the generator of special conformal transformation is a second order differential operator. When applied to the scalar 3-point function, written in this case as a function of the two momenta $p_f$ and $p_i$ only, it gives the Ward identity
\begin{equation}
	\widehat{K}^\mu \llangle \phi_f(p_f) \phi_0(-p_f - p_i) \phi_i(p_i) \rrangle = 0,
\label{eq:K:operator}
\end{equation}
where 
\begin{equation}
	\widehat{K}^\mu \equiv \sum_{p \in \{p_i, p_f\} } 
	\left[ -2 p^\rho \frac{\partial^2}{\partial p^\mu \partial p^\rho}
	+ p^\mu \frac{\partial^2}{\partial p^\rho \partial p^\rho}
	+ 2 (\Delta - d) \frac{\partial}{\partial p^\mu} \right].
\label{eq:K:definition}
\end{equation}
This equation is a Lorentz vector with $d$ components. However, its projection along a direction perpendicular to both $p_i$ and $p_f$ is trivial. There are thus only 2 scalar equations that are generated by $p_f \cdot \widehat{K}$ and $p_i \cdot \widehat{K}$. Using the ansatz \eqref{eq:scalar3ptfct}, these two equations become partial differential equations for the function $F$,
\begin{equation}
\begin{aligned}
	\bigg[ z_f (1-z_f) \frac{\partial^2}{\partial z_f^2}
	- 2 z_f z_i \frac{\partial^2}{\partial z_f \partial z_i} 
	- z_i^2 \frac{\partial^2}{\partial z_i^2} 
	\qquad\qquad &
	\\
	+ \left( 1 + \frac{d}{2} - \Delta_f + \alpha z_f \right) \frac{\partial}{\partial z_f}
	+ \alpha z_i \frac{\partial}{\partial z_i} 
	- \beta
	& \bigg] F(z_f, z_i) = 0,
	\\
	\bigg[ z_i (1-z_i) \frac{\partial^2}{\partial z_i^2}
	- 2 z_f z_i \frac{\partial^2}{\partial z_f \partial z_i} 
	- z_f^2 \frac{\partial^2}{\partial z_f^2} 
	\qquad\qquad &
	\\
	+ \left( 1 + \frac{d}{2} - \Delta_i + \alpha z_i \right) \frac{\partial}{\partial z_i}
	+ \alpha z_f \frac{\partial}{\partial z_f} 
	- \beta
	& \bigg] F(z_f, z_i) = 0,
\end{aligned}
\label{eq:conformalWardidentities}
\end{equation}
with
\begin{equation}
	\alpha = \Delta_f + \Delta_i - \frac{3d}{2} - 1,
	\qquad
	\beta = \frac{\left( \Delta_f + \Delta_0 + \Delta_i - 2d \right)
	\left( \Delta_f - \Delta_0 + \Delta_i - d \right)}{4}.
\end{equation}
This system of equation is of the type satisfied by Appell's $F_4$ generalized hypergeometric function of two variables~\cite[eq.~16.14.4]{NIST:DLMF}. The $F_4$ function is defined around $(z_f, z_i) = (0,0)$ by the double infinite series
\begin{equation}
	F_4( a, b; c_f, c_i; z_f, z_i)
	= \sum_{n,m =0}^\infty
	\frac{(a)_{n+m} (b)_{n+m}}{n! m! (c_f)_n (c_i)_m} \, z_f^n z_i^m.
	\label{eq:AppellF4}
\end{equation}
The most general solution to the system \eqref{eq:conformalWardidentities} is a linear combination of the four functions
\begin{equation}
\begin{array}{c@{\qquad\quad}c}
	(-z_f)^{\Delta_f - d/2} (-z_i)^{\Delta_i - d/2} F_{\Delta_f \Delta_0 \Delta_i}(z_f, z_i),
	& (-z_i)^{\Delta_i - d/2} F_{\Deltatilde_f \Delta_0 \Delta_i}(z_f, z_i),
	\\
	(-z_f)^{\Delta_f - d/2} F_{\Delta_f \Delta_0 \Deltatilde_i}(z_f, z_i),
	& F_{\Deltatilde_f \Delta_0 \Deltatilde_i}(z_f, z_i),
\end{array}
\label{eq:4solutions}
\end{equation}
where we have introduced a shorthand notation for the Appell $F_4$ function
\begin{equation}
	F_{\Delta_f \Delta_0 \Delta_i}(z_f, z_i)
	= F_4\left( \tfrac{\Delta_f - \Delta_0 + \Delta_i}{2},
	\tfrac{\Delta_f - \Deltatilde_0 + \Delta_i}{2};
	\Delta_f - \tfrac{d-2}{2},
	\Delta_i - \tfrac{d-2}{2};
	z_f, z_i \right)
\label{eq:F}
\end{equation}
and denoted $\Deltatilde = d - \Delta$.%
\footnote{This notation is of course reminiscent of the scaling dimension of ``shadow'' operators (see e.g.~ref.~\cite{SimmonsDuffin:2012uy} for a modern discussion). The fact that the Ward identity has four solutions is actually related to the existence of a shadow transform that can be applied either to the initial or to the final state, or to both.
The 3 discarded solutions correspond then to the correlation function $\llangle \widetilde{\phi}_f \phi_0 \phi_i \rrangle$, $\llangle \phi_f \phi_0 \widetilde{\phi}_i \rrangle$ and  $\llangle \widetilde{\phi}_f \phi_0 \widetilde{\phi}_i \rrangle$. Note that in a Wightman function it is not possible to define the shadow transform of the middle operator $\phi_0$. This is reflected in the fact that the solutions are invariant under $\Delta_0 \leftrightarrow d - \Delta_0$, i.e.~$F_{\Delta_f \Delta_0 \Delta_i}(z_f, z_i) = F_{\Delta_f \Deltatilde_0 \Delta_i}(z_f, z_i)$.}

Of the four solutions \eqref{eq:4solutions}, only the first one is consistent with the asymptotic behavior~\eqref{eq:limit:massless:F} for generic values of the scaling dimensions $\Delta_f$ and $\Delta_i$. 
We conclude that 
\begin{equation}
	\llangle \phi_f(p_f) \phi_0(p_0) \phi_i(p_i) \rrangle
	= \lambdatilde_{f0i}
	\Theta(-p_f) \Theta(p_i)
	\frac{(-p_f^2)^{\Delta_f - d/2} (-p_i^2)^{\Delta_i - d/2} }{(p_0^2)^{(\Delta_i + \Delta_f - \Delta_0)/2}}
	F_{\Delta_f \Delta_0 \Delta_i}\left( \frac{p_f^2}{p_0^2}, \frac{p_i^2}{p_0^2} \right)
\label{eq:3scalars:spacelike}
\end{equation}
with an unknown coefficient $\lambdatilde_{f0i}$. This is the equation quoted in the introduction and can be considered the main result of this work. Its simplicity is striking when compared with the integral representation in terms of Bessel functions of ref.~\cite{Bautista:2019qxj}.
In the special case where $\Delta_i, \Delta_f = \frac{d}{2} + n$  with $n \in \mathbb{N}$, the OPE limit~\eqref{eq:limit:massless} does not unambiguously select a unique solution. This situation can however be understood by analytic continuation in $\Delta_i$ and $\Delta_f$ of the general case, since a careful analysis of the Fourier transform shows that Wightman functions do not have non-analyticities when $\Delta = \frac{d}{2} + n$ \cite{Bautista:2019qxj}.
The result~\eqref{eq:3scalars:spacelike} is indeed analytic in $\Delta_f$, $\Delta_0$, $\Delta_i$ and $d$ as long as the unitarity bound is satisfied ($\Delta \geq \frac{d-2}{2}$), and it applies therefore in all generality.

This result is not complete, however, because some kinematically-allowed range of the arguments $p_f^2/p_0^2$ and $p_i^2/p_0^2$ fall outside the radius of convergence of the Appell $F_4$ series. In particular, there is a singularity as $p_0^2 \to 0$ that need to be resolved. This will be done in the next section with the help of the other momentum-space OPE limit.

\subsection{Analytic continuation and normalization}
\label{subsec:analyticcontinuation}

At fixed $z_f$, the radius of convergence of the double hypergeometric series~\eqref{eq:AppellF4} in $z_i$ is $( 1 - \sqrt{|z_f|} )^2$. However, the first singularity in $z_i$ appears on the positive real axis, and it turns out that the Appell $F_4$ function is analytic over the full negative real axis $z_i \in (-\infty, 0]$.
This is made manifest by the existence of a transformation formula stating that~\cite[eq.~16.16.10]{NIST:DLMF}
\begin{equation}
\begin{aligned}
	F_{\Delta_f \Delta_0 \Delta_i}(z_f, z_i)
	&= c_{\Delta_f \Delta_0 \Delta_i}
	(-z_i)^{-(\Delta_f + \Delta_i - \Deltatilde_0)/2}
	F_{\Delta_f \Delta_i \Delta_0}\left( \frac{z_f}{z_i}; \frac{1}{z_i} \right)
	\\
	& \quad
	+ c_{\Delta_f \Deltatilde_0 \Delta_i} 
	(-z_i)^{-(\Delta_f + \Delta_i - \Delta_0)/2}
	F_{\Delta_f \Delta_i \Deltatilde_0}\left( \frac{z_f}{z_i}; \frac{1}{z_i} \right),
\end{aligned}
\label{eq:F4transformation}
\end{equation}
where
\begin{equation}
	c_{\Delta_f \Delta_0 \Delta_i} = \frac{\Gamma\left( \Delta_i - \frac{d-2}{2} \right)
	\Gamma\left( \frac{d}{2} - \Delta_0 \right)}
	{\Gamma\left( \frac{\Delta_f + \Delta_i - \Delta_0}{2} \right)
	\Gamma\left( 1 - \frac{\Delta_f + \Delta_0 - \Delta_i}{2} \right)}.
\label{eq:c}
\end{equation}
For the Wightman 3-point function, this transformation formula and its conjugate taking $(z_f, z_i) \to (1/z_f, z_i/z_f)$
show that the result~\eqref{eq:3scalars:spacelike} applies over the whole kinematic range in which $p_0$ is space-like.
When $p_0$ approaches the light cone, there is a branch point singularity: applying the transformation \eqref{eq:F4transformation} to the 3-point function, 
\begin{equation}
\begin{aligned}
	\llangle \phi_f(p_f) \phi_0(p_0) \phi_i(p_i) \rrangle \qquad &
	\\
	= \lambdatilde_{f0i} \Theta(-p_f) \Theta(p_i)\bigg[ &
	c_{\Delta_f \Deltatilde_0 \Delta_i}
	(-p_f^2)^{\Delta_f - d/2} (-p_i^2)^{(\Delta_i + \Delta_0 - \Delta_f - d)/2}
	F_{\Delta_f \Delta_i \Deltatilde_0}\left( \frac{p_f^2}{p_i^2}, \frac{p_0^2}{p_i^2} \right)
	\\
	& + c_{\Delta_f \Delta_0 \Delta_i}
	\frac{(-p_f^2)^{\Delta_f - d/2} (p_0^2)^{\Delta_0 - d/2}}{(-p_i^2)^{(\Delta_f + \Delta_0 - \Delta_i)/2}}
	F_{\Delta_f \Delta_i \Delta_0}\left( \frac{p_f^2}{p_i^2}, \frac{p_0^2}{p_i^2} \right) \bigg].
\end{aligned}
\label{eq:3scalars:analyticcontinuation}
\end{equation}
The first term on the right-hand side is analytic at $p_0^2 = 0$, and the non-analyticity only arises from the factor $(p_0^2)^{\Delta_0 - d/2}$ in the second term.%
\footnote{The divergence present when $\Delta_0 < \frac{d}{2}$ is actually integrable by the unitarity bound $\Delta_0 > \frac{d-2}{2}$. This suggests that it should be possible to take the inverse Fourier transform of this expression and recover the position-space 3-point function.}
This representation suggests that the 3-point function can be continued past the light cone $p_0^2 = 0$, but the continuation is ambiguous.

In fact, the structure of eq.~\eqref{eq:3scalars:analyticcontinuation} is not surprising: when expressing the conformal Ward identities in terms of the variables $p_f^2/p_i^2$ and $p_0^2/p_i^2$, they still admit four solutions, of which only two are consistent with the asymptotic limit \eqref{eq:limit:pf2:zero}. The most general form of the 3-point function at time-like $p_0$ is therefore
\begin{equation}
\begin{aligned}
	\llangle \phi_f(p_f) \phi_0(p_0) \phi_i(p_i) \rrangle \qquad &
	\\
	= \Theta(-p_f) \Theta(p_i)\bigg[ &
	\lambdatilde_{f0i}^{(a)} \, 
	(-p_f^2)^{\Delta_f - d/2} (-p_i^2)^{(\Delta_i + \Delta_0 - \Delta_f - d)/2}
	F_{\Delta_f \Delta_i \Deltatilde_0}\left( \frac{p_f^2}{p_i^2}, \frac{p_0^2}{p_i^2} \right)
	\\
	& + \lambdatilde_{f0i}^{(b)} \, 
	\frac{(-p_f^2)^{\Delta_f - d/2} (-p_0^2)^{\Delta_0 - d/2}}{(-p_i^2)^{(\Delta_f + \Delta_0 - \Delta_i)/2}}
	F_{\Delta_f \Delta_i \Delta_0}\left( \frac{p_f^2}{p_i^2}, \frac{p_0^2}{p_i^2} \right) \bigg].
\end{aligned}
\label{eq:3scalars:timelike}
\end{equation}
The two unknown coefficients $\lambdatilde_{f0i}^{(a)}$ and $\lambdatilde_{f0i}^{(b)}$ can be fixed with the help of the OPE limit $p_f \to 0$.  This limit is subtle, however, since the second argument of the Appell functions in eq.~\eqref{eq:3scalars:timelike} approaches their radius of convergence, $p_0^2 / p_i^2 \to 1$, and it is not possible to evaluate them using the hypergeometric series definition \eqref{eq:AppellF4}.
Instead, one can study the OPE limit by taking $p_f^2 \to 0_-$ first, as shown in figure~\eqref{fig:kinematics:limits}~(a). In this case, the Appell functions turn into ordinary hypergeometric functions
\begin{equation}
	F_{\Delta_f \Delta_i \Delta_0}(0, z)
	= \, _2F_1\left( \tfrac{\Delta_f - \Delta_i + \Delta_0}{2},
	\tfrac{\Delta_f - \Deltatilde_i + \Delta_0}{2};
	\Delta_0 - \tfrac{d-2}{2}; z \right),
\end{equation}
with a well-known behavior at argument $z = 1$, where we have the asymptotic limit
\begin{equation}
\begin{aligned}
	F_{\Delta_f \Delta_i \Delta_0}(0, z)
	&\approx \frac{\Gamma\left( \Delta_0 - \frac{d-2}{2} \right)
	\Gamma\left( 1 - \Delta_f \right)}
	{\Gamma\left( 1 - \frac{\Delta_f + \Delta_i - \Delta_0}{2} \right)
	\Gamma\left( 1 - \frac{\Delta_f + \Deltatilde_i - \Delta_0}{2} \right)}
	\left[ 1 + \O(1-z) \right]
	\\
	& \quad
	+ \frac{\Gamma\left( \Delta_0 - \frac{d-2}{2} \right)
	\Gamma\left( \Delta_f - 1 \right)}
	{\Gamma\left( \frac{\Delta_f - \Delta_i + \Delta_0}{2} \right)
	\Gamma\left( \frac{\Delta_f - \Deltatilde_i + \Delta_0}{2} \right)}
	(1-z)^{1 - \Delta_f}
	\left[ 1 + \O(1-z) \right].
\end{aligned}
\label{eq:F:limit:pf:zero}
\end{equation}
The non-analytic term in $1-z$ in the second line is in contradiction with the existence of the limit $p_f \to 0$. Therefore, the non-analytic terms coming from the functions $F_{\Delta_f \Delta_i \Delta_0}$ and $F_{\Delta_f \Delta_i \Deltatilde_0}$ in eq.~\eqref{eq:3scalars:timelike} must cancel exactly, and the analytic terms must add up to the limit determined in eq.~\eqref{eq:limit:pf:zero}.
This gives a linear system of equations for the coefficients $\lambdatilde_{f0i}^{(a)}$ and $\lambdatilde_{f0i}^{(b)}$, whose unique solution is
\begin{equation}
	\lambdatilde_{f0i}^{(a)}
	= \frac{\Gamma\left( \Delta_0 - \frac{d}{2} \right)}
	{\Gamma\left( \frac{\Delta_f + \Delta_i - \Deltatilde_0}{2} \right)
	\Gamma\left( \frac{\Delta_f + \Delta_0 - \Delta_i}{2} \right)} \, \mathcal{N},
	\qquad
	\lambdatilde_{f0i}^{(b)}
	= \frac{\Gamma\left( \frac{d}{2} - \Delta_0 \right)}
	{\Gamma\left( \frac{\Delta_f + \Delta_i - \Delta_0}{2} \right)
	\Gamma\left( \frac{\Delta_f + \Deltatilde_0 - \Delta_i}{2} \right)} \, \mathcal{N},
\label{eq:lambdatilde:ab}
\end{equation}
where $\mathcal{N}$ is related to the OPE coefficient $\lambda_{f0i}$ by
\begin{equation}
	\mathcal{N} = \frac{2^{2d - \Delta_f - \Delta_0 - \Delta_i + 2} \pi^{d+2}}
	{\Gamma\left( \frac{\Delta_i + \Delta_0 - \Delta_f}{2} \right)
	\Gamma\left( \Delta_f - \frac{d-2}{2} \right)
	\Gamma\left( \frac{\Delta_i + \Delta_0 - \Delta_f - d + 2}{2} \right)} \, \lambda_{f0i}.
\end{equation}
This result is valid for any configuration of momenta as long as $p_0$ is space-like and $p_f^2 < p_i^2$.
There exists a similar expression covering the case $p_f^2 > p_i^2$, which by the conjugation symmetry \eqref{eq:scalar3pt:conjugation} can be obtained from the simultaneous exchange $p_i^2 \leftrightarrow p_f^2$ and $\Delta_i \leftrightarrow \Delta_f$ in eq.~\eqref{eq:3scalars:timelike}.
The special case $p_i^2 = p_f^2$ is covered by eq.~\eqref{eq:3scalars:spacelike} since it necessarily implies that $p_0$ is space-like.

The result \eqref{eq:3scalars:timelike} also turns out to be a straightforward analytic continuation of eq.~\eqref{eq:3scalars:analyticcontinuation}, sharing the same structure with a branch point singularity at $p_0^2 = 0$. The parts that are analytic in $p_0^2$ match provided that one makes the identification
\begin{equation}
	\lambdatilde_{f0i} = \frac{2^{2d - \Delta_f - \Delta_0 - \Delta_i + 2} \pi^{d+2}}
	{\Gamma\left( \frac{\Delta_f + \Delta_0 - \Delta_i}{2} \right)
	\Gamma\left( \frac{\Delta_i + \Delta_0 - \Delta_f}{2} \right)
	\Gamma\left( \Delta_f - \frac{d-2}{2} \right)
	\Gamma\left( \Delta_i - \frac{d-2}{2} \right)}
	\lambda_{f0i}.
\label{eq:lambdatilde}
\end{equation}
$\lambdatilde_{f0i}$ is real and analytic in the scaling dimensions $\Delta_f$, $\Delta_0$ and $\Delta_i$ as well as in the space-time dimension $d$. It has zeroes when $\Delta_f = \Delta_i + \Delta_0 + 2n$ and when $\Delta_i = \Delta_f + \Delta_0 + 2n$ with $n \in \mathbb{N}$. 
This is the situation of generalized free field theory discussed in more detail in section~\ref{subsec:GFF}.
In contrast, the 3-point function \eqref{eq:3scalars:timelike} at time-like $p_0$ does not vanish in generalized free field theory, but it it still analytic in the scaling dimensions, even though the coefficients $\lambdatilde_{f0i}^{(a)}$ and $\lambdatilde_{f0i}^{(b)}$ are not. This is best seen from the following compact expression for the Wightman 3-point function that covers both the space-like and the time-like regions in $p_0$:
\begin{equation}
\begin{aligned}
	& \llangle \phi_f (p_f) \phi_0(p_0) \phi_i(p_i) \rrangle
	\\
	& \qquad
	= \lambdatilde_{f0i} \Theta(-p_f) \Theta(p_i)
	\frac{(-p_f^2)^{\Delta_f - d/2} (-p_i^2)^{\Delta_i - d/2}}{(p_0^2 - i \epsilon)^{(\Delta_f + \Delta_i - \Delta_0)/2}}
	F_{\Delta_f \Delta_0 \Delta_i}\left( \frac{p_f^2}{p_0^2 - i \epsilon}, \frac{p_i^2}{p_0^2 - i \epsilon} \right)
	\\
	& \qquad\quad\qquad
	+ \lambdatilde_{fi0} \Theta(-p_f) \Theta(-p_0)
	\frac{(-p_f^2)^{\Delta_f - d/2} (-p_0^2)^{\Delta_0 - d/2}}{(p_i^2 + i \epsilon)^{(\Delta_f + \Delta_0 - \Delta_i)/2}}
	F_{\Delta_f \Delta_i \Delta_0}\left( \frac{p_f^2}{p_i^2}, \frac{p_0^2}{p_i^2} \right)
	\\
	& \qquad\quad\qquad
	+ \lambdatilde_{0fi} \Theta(p_i)\Theta(p_0)
	\frac{(-p_i^2)^{\Delta_i - d/2} (-p_0^2)^{\Delta_0 - d/2}}{(p_f^2 + i \epsilon)^{(\Delta_i + \Delta_0 - \Delta_f)/2}}
	F_{\Delta_0 \Delta_f \Delta_i}\left( \frac{p_0^2}{p_f^2}, \frac{p_i^2}{p_f^2} \right).
\end{aligned}
\label{eq:3scalars:complete}
\end{equation}
This representation somehow obscures the facts that the 3-point function is real and that it has a branch point at $p_0^2 = 0$, but it makes the analyticity in $\Delta_f$, $\Delta_0$, $\Delta_i$ and $d$ manifest over its whole region of support.

This is our final result for the Wightman 3-point function of scalar operators.
It should be noted that this result has been successfully compared with a direct evaluation of the Fourier transform of the position-space 3-point function, analytically in the OPE limits of section~\ref{subsec:OPElimits} and numerically for generic kinematics. Some details about the direct computation of the Fourier transform are presented in appendix~\ref{app:Fourierintegrals}.
Before moving on to the study of other correlation functions, we will discuss some interesting features of the Wightman function.

\subsection{Generalized free field theory}
\label{subsec:GFF}

As already mentioned, something special happens to the 3-point function when one scaling dimension equals the sum of the other two.
Let us assume that $\Delta_0 = \Delta_f + \Delta_i$ first. In this case we will interpret $\phi_0$ as the composite operator $[\phi_f \phi_i]$. Of the three terms in eq.~\eqref{eq:3scalars:complete}, only the first one remains because $\lambdatilde_{fi0} = \lambdatilde_{0fi} = 0$. Moreover, the Appell function $F_{\Delta_f\Delta_0\Delta_i}$ is trivially equal to one in this case. We obtain therefore
\begin{equation}
	\llangle \phi_f(p_f) [\phi_f \phi_i](p_0) \phi_i(p_i) \rrangle
	= \lambdatilde_{\phi_f[\phi_f\phi_i]\phi_i} \Theta(-p_f) \Theta(p_i)
	(-p_f^2)^{\Delta_f - d/2} (-p_i^2)^{\Delta_i - d/2}.
\end{equation}
The dependence on the momenta $p_i$ and $p_f$ factorizes, and we see a similar factorization in the OPE coefficient,
\begin{equation}
	\lambdatilde_{\phi_f[\phi_f\phi_i]\phi_i}
	= \left( \frac{2^{d- 2 \Delta_f + 1} \pi^{(d+2)/2}}
	{\Gamma\left( \Delta_f \right) \Gamma\left( \Delta_f - \frac{d-2}{2} \right)} \right)
	\left( \frac{2^{d- 2 \Delta_i + 1} \pi^{(d+2)/2}}
	{\Gamma\left( \Delta_i \right) \Gamma\left( \Delta_i - \frac{d-2}{2} \right)} \right)
	\lambda_{\phi_f[\phi_f\phi_i]\phi_i}.
\end{equation}
This means that we can write
\begin{equation}
	\llangle \phi_f(p_f) [\phi_f \phi_i](p_0) \phi_i(p_i) \rrangle
	= \lambda_{\phi_f[\phi_f\phi_i]\phi_i}
	\llangle \phi_f(p_f) \phi_f(-p_f) \rrangle
	\llangle \phi_i(-p_i) \phi_i(p_i) \rrangle.
\label{eq:doubletrace:fi}
\end{equation}
This result is expected from a generalized free field theory, and since the only dynamical data in a conformal 3-point function is encoded in the OPE coefficient, the kinematics must match that of the generalized free theory whenever the scaling dimensions obey such relations.

A similar study of the case $\Delta_i = \Delta_f + \Delta_0$ shows that 
\begin{equation}
	\llangle \phi_f(p_f) \phi_0(p_0) [\phi_f \phi_0](p_i) \rrangle
	= \lambda_{\phi_f\phi_0[\phi_f\phi_0]}
	\llangle \phi_f(p_f) \phi_f(-p_f) \rrangle
	\llangle \phi_0(p_0) \phi_0(-p_0) \rrangle.
\label{eq:doubletrace:f0}
\end{equation}
If instead we take $\Delta_0 = \Delta_f + \Delta_i + 2n$ where $n$ is a positive integer, the right-hand side of eq.~\eqref{eq:doubletrace:fi} gets multiplied by a homogeneous polynomial of degree $n$ in $p_f^2$, $p_i^2$ and $p_f \cdot p_i$, because the hypergeometric series that defines $F_{\Delta_f \Delta_0 \Delta_i}$ terminates at order $n$.
This provides a way of resolving the exact structure of the double-trace operator $[ \phi_f \square^n \phi_i]$.

The factorization of 3-point functions into 2-point functions is actually a trivial statement in the position-space representation, and it is easy to take their Fourier transform directly and reproduce expressions like \eqref{eq:doubletrace:fi} and \eqref{eq:doubletrace:f0}.
Nevertheless, it is important to see that our general result \eqref{eq:3scalars:complete} covers these special cases in a quite non-trivial manner.

\subsection{Holomorphic factorization in two dimensions}
\label{subsec:holomorphicfactorization}

Another curiosity occurs in two space-time dimension. Using light-cone coordinates $p^2 = -p^+ p^-$ together with a special identity of the Appell $F_4$ function that only applies when $d = 2$~\cite[eq.~16.16.6]{NIST:DLMF}, we can write
\begin{equation}
\begin{aligned}
	F_{\Delta_f \Delta_0 \Delta_i}\left( \frac{p_f^2}{p_0^2}, \frac{p_i^2}{p_0^2} \right)
	&= \, _2F_1\left( \tfrac{\Delta_f - \Delta_0 + \Delta_i}{2},
	\tfrac{\Delta_f - \Deltatilde_0 + \Delta_i}{2}; \Delta_f;
	- \frac{p_f^+}{p_0^+} \right)
	\\
	& \quad \times
	\, _2F_1\left( \tfrac{\Delta_f - \Delta_0 + \Delta_i}{2},
	\tfrac{\Delta_f - \Deltatilde_0 + \Delta_i}{2}; \Delta_i;
	-\frac{p_i^-}{p_0^-} \right).
\end{aligned}
\end{equation}
This allows to write the 3-point function in the fully factorized form
\begin{equation}
	\llangle \phi_f(p_f) \phi_0(p_0) \phi_i(p_i) \rrangle 
	= \lambda_{f0i} \, W(p_f^+, p_0^+, p_i^+)  W(p_f^-, p_0^-, p_i^-)
\end{equation}
where we have defined
\begin{equation}
\begin{aligned}
	W(p_f^+, p_0^+, p_i^+)
	&= \frac{(2\pi)^2}{2^{(\Delta_f + \Delta_0 + \Delta_i -2)/2}}
	\frac{(p_f^+)^{\Delta_f - d/2} (p_i^+)^{\Delta_i - d/2}}{|p_0^+|^{(\Delta_f - \Delta_0 + \Delta_i)/2}}
	\\
	& \quad \times
	\Bigg[ \frac{\Theta(p_0^+) }{\Gamma\left( \Delta_i \right)
	\Gamma\left( \frac{\Delta_f + \Delta_0 - \Delta_i}{2} \right)}
	\, _2F_1\left( \tfrac{\Delta_f - \Delta_0 + \Delta_i}{2},
	\tfrac{\Delta_f - \Deltatilde_0 + \Delta_i}{2}; \Delta_i;
	-\frac{p_i^+}{p_0^+} \right)
	\\
	& \quad \quad
	+ \frac{\Theta(-p_0^+) }{\Gamma\left( \Delta_f \right)
	\Gamma\left( \frac{\Delta_i + \Delta_0 - \Delta_f}{2} \right)}
	\, _2F_1\left( \tfrac{\Delta_f - \Delta_0 + \Delta_i}{2},
	\tfrac{\Delta_f - \Deltatilde_0 + \Delta_i}{2}; \Delta_f;
	- \frac{p_f^+}{p_0^+} \right) \bigg].
\end{aligned}
\end{equation}
$\Theta$ here is the ordinary Heaviside step function.
This is consistent with the fact that the Wightman 3-point function in position space can be factorized into holomorphic and anti-holomorphic pieces, or equivalently into left- and right-movers. This result is also found to match the direct Fourier transform of the position-space correlator, which can be easily performed in this case.
The interesting way in which this factorization arises from the general expression~\eqref{eq:3scalars:complete} is another verification of its validity.


\section{Adding spin: traceless symmetric tensor}
\label{sec:spinning}

We will now discuss how to incorporate an operator that is not a scalar in the analysis of the previous section.
Our approach is not meant to be systematic, but instead focuses on the simplest case as an example.

\subsection{Poincar\'e and scale symmetry}

Starting with the Wightman 3-point function~\eqref{eq:mainresult}, we choose to keep the operators $\phi_0$ and $\phi_i$ scalar and replace $\phi_f$ by an operator $\O_f$ carrying spin. The only type of spin representations allowed by conformal symmetry are are traceless symmetric tensors.
In this case it is convenient to introduce a null polarization vector $\zeta^2 = 0$ and define the momentum-space operator with spin $\ell$ by~\cite{Costa:2011mg, Dolan:2011dv, Costa:2016hju}
\begin{equation}
	\O^{(\ell)}(p, \zeta) = \zeta^{\mu_1} \cdots \zeta^{\mu_\ell} \O^{\mu_1 \ldots \mu_\ell}(p).
\label{eq:spinningoperator}
\end{equation}
Both the symmetry and the tracelessness of the operator are automatically encoded in this definition.
As in the scalar case, this operator is in one-to-one correspondence with a momentum eigenstate $\ket{ \O^{(\ell)}(p, \zeta) } \equiv \O^{(\ell)}(p, \zeta) \ket{0}$. The only novelty is that not all such state are linearly independent, since states related by a little group transformation on $\zeta$ are equivalent, and some states are even null in the case of a conserved operator $\partial_\mu \O^{\mu\nu\ldots} = 0$. But these considerations do not affect the construction of section~\ref{subsec:momentumeigenstates}. We can still construct the most general ansatz consistent with Poincar\'e and scale symmetry, the only new constraint being that it must be a polynomial of degree $\ell$ in the polarization vector $\zeta$.
Therefore we can write
\begin{equation}
\begin{aligned}
	\llangle \O_f^{(\ell)}(p_f, \zeta) \phi_0(p_0) \phi_i(p_i) \rrangle
	&= \Theta(-p_f) \Theta(p_i)
	(p_0^2)^{(\Delta_f + \Delta_0 + \Delta_i - \ell - 2d)/2}
	\\
	& \quad \times
	\sum_{n = 0}^\ell
	(p_f \cdot \zeta)^n (p_i \cdot \zeta)^{\ell - n}
	F^{(\ell)}_n\left( \frac{p_f^2}{p_0^2}, \frac{p_i^2}{p_0^2} \right)
\end{aligned}
\label{eq:ansatz:spinning}
\end{equation}
where the $F^{(\ell)}_n$ are $\ell$ distinct functions to be determined. This ansatz is valid as long as $p_0$ is space-like as in figure~\ref{fig:kinematics}~(a), and the general case will again be obtained by analytic continuation.

Note that this treatment of the spin does not apply in $d = 2$ spacetime dimensions:
In that case the ansatz \eqref{eq:ansatz:spinning} is redundant because the polarization vector $\zeta$ can be expressed as a linear combination of $p_f$ and $p_i$.
In $d = 2$, all operators can be viewed as scalars with different conformal weights for the holomorphic and anti-holomorphic pieces, and it is easy to generalize the results of section~\ref{subsec:holomorphicfactorization} in that case.

\subsection{Conformal Ward identities}

Further restrictions on the functions $F^{(\ell)}_n$ in eq.~\eqref{eq:ansatz:spinning} are provided by the Ward identity for special conformal transformations
\begin{equation}
	\widehat{K}^\mu \llangle \O_f^{(\ell)}(p_f, \zeta) \phi_0(p_0) \phi_i(p_i) \rrangle = 0,
\label{eq:K:operator:spinning}
\end{equation}
where now instead of eq.~\eqref{eq:K:definition} the differential operator is
\begin{equation}
	\widehat{K}^\mu \equiv \sum_{p \in \{p_i, p_f\} } 
	\left[ -2 p^\rho \frac{\partial^2}{\partial p^\mu \partial p^\rho}
	+ p^\mu \frac{\partial^2}{\partial p^\rho \partial p^\rho}
	+ 2 (\Delta - d) \frac{\partial}{\partial p^\mu} \right]
	+ \frac{\partial}{\partial p_f^\rho}
	\left(  \zeta^\mu \frac{\partial}{\partial \zeta^\rho}
	- \zeta^\rho \frac{\partial}{\partial \zeta^\mu} \right).
\label{eq:K:definition:spinning}
\end{equation}
The ansatz~\eqref{eq:ansatz:spinning} contains more freedom than its scalar counterpart eq.~\eqref{eq:scalar3ptfct} as it is written in terms of $\ell$ distinct unknown functions, but it should be noted that the Ward identity~\eqref{eq:K:operator:spinning} is also more constraining than eq.~\eqref{eq:K:operator}: it does not only have components in the plane spanned by $p_i$ and $p_f$, but also along the orthogonal direction.
If we denote by $p_\perp$ a vector such that $p_\perp \cdot p_i = p_\perp \cdot p_f = 0$, then the Ward identity generated by the operator $p_\perp \cdot \widehat{K}$ takes the form
\begin{equation}
\begin{aligned}
	& (n + 1) ( \Delta_f - 2 + \ell - n ) F^{(\ell)}_{n+1}(z_f, z_i)
	\\
	& \qquad
	= (\ell - n) \bigg[
	\Delta_i - d + 1 - \ell + n
	+ (1 - z_f - z_i) \frac{\partial}{\partial z_f}
	- 2 z_i \frac{\partial}{\partial z_i} \bigg] F^{(\ell)}_n(z_f, z_i).
\end{aligned}
\label{eq:Fn:recursion}
\end{equation}
This recursion relation determines all the functions $F^{(\ell)}_n(z_f, z_i)$ in terms of $F^{(\ell)}_0(z_f, z_i)$. Moreover the projections of the differential operator \eqref{eq:K:definition:spinning} along the direction of $p_f$ and $p_i$ are such that they never raise the power of $p_i \cdot \zeta$, which means that we get a closed system of differential equations for $F^{(\ell)}_0(z_f, z_i)$, which reads
\begin{equation}
\begin{aligned}
	\bigg[ z_f (1-z_f) \frac{\partial^2}{\partial z_f^2}
	- 2 z_f z_i \frac{\partial^2}{\partial z_f \partial z_i} 
	- z_i^2 \frac{\partial^2}{\partial z_i^2} 
	\qquad\qquad &
	\\
	+ \left( 1 + \frac{d}{2} - \Delta_f + \alpha z_f \right) \frac{\partial}{\partial z_f}
	+ \alpha z_i \frac{\partial}{\partial z_i} 
	- \beta
	& \bigg] F^{(\ell)}_0(z_f, z_i) = 0,
	\\
	\bigg[ z_i (1-z_i) \frac{\partial^2}{\partial z_i^2}
	- 2 z_f z_i \frac{\partial^2}{\partial z_f \partial z_i} 
	- z_f^2 \frac{\partial^2}{\partial z_f^2} 
	\qquad\qquad &
	\\
	+ \left( 1 + \frac{d}{2} - \Delta_i + \alpha z_i \right) \frac{\partial}{\partial z_i}
	+ \alpha z_f \frac{\partial}{\partial z_f} 
	- \beta
	& \bigg] F^{(\ell)}_0(z_f, z_i) = 0,
\end{aligned}
\end{equation}
with
\begin{equation}
	\alpha = \Delta_f + \Delta_i - \frac{3d}{2} - 1 + \ell,
	\qquad
	\beta = \frac{\left( \Delta_f + \Delta_0 + \Delta_i - \ell - 2d \right)
	\left( \Delta_f - \Delta_0 + \Delta_i - d - \ell \right)}{4}.
\end{equation}
This system is identical to that of eq.~\eqref{eq:conformalWardidentities} but with different parameters $\alpha$ and $\beta$.
Hence its most general solution is also a linear combinations of four Appell $F_4$ hypergeometric functions,
\begin{equation}
\begin{array}{c@{\qquad\quad}c}
	(-z_f)^{\Delta_f - d/2} (-z_i)^{\Delta_i - d/2} F^{(\ell)}_{\Delta_f \Delta_0 \Delta_i}(z_f, z_i),
	& (-z_i)^{\Delta_i - d/2} F^{(\ell)}_{\Deltatilde_f \Delta_0 \Delta_i}(z_f, z_i),
	\\
	(-z_f)^{\Delta_f - d/2} F^{(\ell)}_{\Delta_f \Delta_0 \Deltatilde_i}(z_f, z_i),
	& F^{(\ell)}_{\Deltatilde_f \Delta_0 \Deltatilde_i}(z_f, z_i),
\end{array}
\label{eq:4solutions:spinning}
\end{equation}
where now
\begin{equation}
	F^{(\ell)}_{\Delta_f \Delta_0 \Delta_i}(z_f, z_i)
	= F_4\left( \tfrac{\Delta_f - \Delta_0 + \Delta_i + \ell}{2},
	\tfrac{\Delta_f - \Deltatilde_0 + \Delta_i + \ell}{2};
	\Delta_f - \tfrac{d-2}{2},
	\Delta_i - \tfrac{d-2}{2};
	z_f, z_i \right).
\label{eq:F:spinning}
\end{equation}
Note that $F^{(\ell)}_{\Delta_f \Delta_0 \Delta_i}(z_f, z_i)$ can be obtained from the scalar function $F_{\Delta_f \Delta_0 \Delta_i}(z_f, z_i)$ by a shift of all scaling dimensions $\Delta \to \Delta + \ell$ accompanied by a shift $d \to d + 2 \ell$ in the space-time dimension, under which the combination $\Delta - \frac{d}{2}$ is invariant.

Without going into the details of it, a logic similar to that of section~\ref{subsec:OPElimits} can be used to argue that among the four solutions $\eqref{eq:4solutions:spinning}$, only the first one is consistent with the OPE.
Thus we arrive at the result
\begin{equation}
	F^{(\ell)}_0(z_f, z_i) = \lambdatilde_{f0i}^{(\ell)} 
	(-z_f)^{\Delta_f - d/2} (-z_i)^{\Delta_i - d/2} F^{(\ell)}_{\Delta_f \Delta_0 \Delta_i}(z_f, z_i),
\label{eq:F0}
\end{equation}
and the other functions $F^{(\ell)}_n(z_f, z_i)$ are defined recursively by eq.~\eqref{eq:Fn:recursion}.%
\footnote{Derivatives of Appell $F_4$ functions can be again expressed in terms of Appell $F_4$ functions with parameters shifted by integers, but we did not find a form simple enough for the generic function $F^{(\ell)}_n(z_f, z_i)$ to be reproduced here.}

\subsection{Analytic continuation and normalization}

The analytic continuation of this result to the regions of time-like $p_0$ proceeds as in section~\ref{subsec:analyticcontinuation}, where we had seen that it is uniquely determined by the existence of the OPE limits $p_f \to 0$ and $p_i \to 0$. 
We do not provide the details of all such analytic continuations here as the result is quite complicated, but focus instead on the simplest case that allows to determine the coefficient $\lambdatilde_{f0i}^{(\ell)}$ in eq.~\eqref{eq:F0}.

Let us study the limit $p_f \to 0$ of the 3-point function. In order to achieve this, we apply the transformation \eqref{eq:F4transformation} to the function $F^{(\ell)}_0$ of eq.~\eqref{eq:F0}, and then continue the non-integer power of $p_0^2$ in such a way that the resulting contribution to the 3-point function is analytic around the point $p_f = 0$. 
When this procedure is complete, we are left with the asymptotic limit
\begin{equation}
\begin{aligned}
	\llangle \O^{(\ell)}_f(p_f, \zeta) \phi_0(p_0) \phi_i(p_i) \rrangle_{p_f \to 0}
	&= \lambdatilde^{(\ell)}_{f0i}
	\frac{\Gamma\left( \Delta_i - \frac{d-2}{2} \right)
	\Gamma\left( \frac{\Delta_f + \Delta_0 - \Delta_i + \ell}{2} \right)}
	{\Gamma\left( \Delta_f + \ell \right)
	\Gamma\left( \frac{\Delta_i + \Delta_0 - \Delta_f - \ell - d + 2}{2} \right)}
	\\
	& \quad \times
	(-p_f^2)^{\Delta_f - d/2} (-p_i^2)^{(\Delta_i + \Delta_0 - \Delta_f - \ell - d)/2}
	\left[ (\zeta \cdot p_i)^\ell + \ldots \right].
\end{aligned}
\label{eq:limit:pi0:spinning}
\end{equation}
The ellipsis indicate terms of order $(\zeta \cdot p_f) / |p_f|$, which arise from the analytic continuation of the functions $F^{(\ell)}_n$ with $n > 0$. It is important to realize that the term $n = 0$ in the ansatz \eqref{eq:ansatz:spinning} is not the only contributor in the limit $p_f \to 0$ since the functions $F^{(\ell)}_n$ contain increasingly divergent powers of $|p_f|$.

This result can be matched with the position-space OPE
\begin{equation}
	\phi_1(x_1) \ket{ \phi_2(x_2) }
	= \lambda_{\O12} \, C_{\O12}^{\mu_1 \ldots \mu_\ell}(x_1 - x_2, P) \ket{ \O^{\mu_1 \ldots \mu_\ell}(x_2) }
	+ \ldots
\end{equation}
where we ignored the contribution of all other operators besides the traceless symmetric tensor $\O^{\mu_1 \ldots \mu_\ell}$.
The operator $C_{\O12}^{\mu_1 \ldots \mu_\ell}(x, P)$ admits a series expansion in $P^\mu$, given at lowest order by
\begin{equation}
	C^{\mu_1 \ldots \mu_\ell}_{\O12}(x, \zeta, P) = \frac{1}{|x|^{\Delta_1 + \Delta_2 - \Delta_\O + \ell}} 
	\left[ x^{\mu_1} \cdots x^{\mu_\ell} + \O(P) \right].
\end{equation}
Taking the Fourier transform of this OPE as in section~\ref{subsec:OPElimits}, we get
\begin{equation}
	\phi_1(p_1) \ket{\phi_2(p_2)}
	= \lambda_{\O12}
	\widetilde{C}^{\mu_1 \ldots \mu_\ell}_{\O12}(p_1, p_1 + p_2) \ket{ \O^{\mu_1 \ldots \mu_\ell}(p_1 + p_2) }
	+ \ldots
\end{equation}
where $\widetilde{C}^{\mu_1 \ldots \mu_\ell}_{\O12}(p, q)$ is the Fourier transform of $C_{\O12}^{\mu_1 \ldots \mu_\ell}(x, q)$, given at lowest order in $q$ by
\begin{equation}
\begin{aligned}
	\widetilde{C}^{\mu_1 \ldots \mu_\ell}_{\O12}(p, 0)
	& = \frac{i^\ell 2^{d- \Delta_1 - \Delta_2 + \Delta_\O + 1} \pi^{(d+2)/2}}
	{\Gamma\left( \frac{\Delta_1 + \Delta_2 - \Delta_\O + \ell}{2} \right)
	\Gamma\left( \frac{\Delta_1 + \Delta_2 - \Delta_\O - \ell - d + 2}{2} \right)}
	\\
	& \quad \times
	(-p^2)^{(\Delta_1 + \Delta_2 - \Delta_\O - \ell - d)/2}
	\left[ p^{\mu_1} \cdots p^{\mu_\ell} + \text{trace terms} \right].
\end{aligned}
\end{equation}
We did not bother to write down the trace terms involving the metric $\eta^{\mu_i \mu_j}$ explicitly as these vanish when contracted with the traceless symmetric tensor $\O^{\mu_1 \ldots \mu_\ell}$.
To use this OPE in the 3-point function, we also need the momentum-space 2-point function of traceless symmetric tensor operators. It was computed for instance in ref.~\cite{Gillioz:2018mto}, and can be written in a compact form as
\begin{equation}
\begin{aligned}
	\llangle \O^{(\ell)}(-p,\zeta) \O^{\mu_1 \ldots \mu_\ell}(p) \rrangle
	&= \frac{2^{d- 2 \Delta + 1} \pi^{(d+2)/2}}
	{(\Delta + \ell - 1) \Gamma( \Delta - 1 )
	\Gamma\left( \Delta - \frac{d-2}{2} \right)}
	\\
	& \quad \times
	\Theta(p) (-p^2)^{\Delta - d/2} \left[ \zeta^{\mu_1} \cdots \zeta^{\mu_\ell}
	+ \ldots \right]
\end{aligned}
\end{equation}
The ellipsis indicate that we have omitted terms proportional to $\zeta \cdot p$.
The structure of these terms is quite complicated, but we do not need them to perform the comparison with eq.~\eqref{eq:limit:pi0:spinning}, where they are neglected as well. We obtain finally
\begin{equation}
	\lambdatilde^{(\ell)}_{f0i}
	= \frac{(-i)^\ell 2^{2d- \Delta_f - \Delta_0 - \Delta_i + 2} \pi^{d+2}
	\left( \Delta_f - 1 \right)_\ell}
	{\Gamma\left( \Delta_f - \frac{d-2}{2} \right)
	\Gamma\left( \Delta_i - \frac{d-2}{2} \right)
	\Gamma\left( \frac{\Delta_f + \Delta_0 - \Delta_i + \ell}{2} \right)
	\Gamma\left( \frac{\Delta_i + \Delta_0 - \Delta_f + \ell}{2} \right)} \,
	\lambda^{(\ell)}_{f0i}.
\end{equation}
Like the coefficient $\lambdatilde_{f0i}$ of eq.~\eqref{eq:lambdatilde}, $\lambdatilde^{(\ell)}_{f0i}$ is analytic in all the scaling dimensions and in $d$, and it has zeroes at the dimensions of double-trace operators, in this case when $\Delta_f = \Delta_i + \Delta_0 + \ell + 2n$ and $\Delta_i = \Delta_f + \Delta_0 + \ell + 2n$ with $n \in \mathbb{N}$. These zeroes are consistent with the vanishing of the 3-point function in generalized free field theory when $p_0$ is space-like. The analytic continuation to time-like $p_0$ comes with a pole that cancels either one of these zeroes, and in that case the Appell $F_4$ function can be expressed as a finite hypergeometric sum. As in the scalar case this can be used to resolve the exact structure of the double-trace operators.

This concludes our study of Wightman functions involving a traceless symmetric tensor. The problem of generalizing our findings to arbitrary spin representations for each of the three operators is left for future work.


\section{Time-ordered products}
\label{sec:timeordering}

In this section we consider correlation function involving time-ordered products of operators and show to what extent the method of section~\ref{sec:scalar} can be used. The results also illustrate how different time-ordered correlation function are from the Wightman function in momentum space.

\subsection{Partial time-ordering}
\label{subsec:partialtimeordering}

We consider first the case in which two out of the three operators in the correlation function are time-ordered, as in
\begin{equation}
	\llangle \phi_f(p_f) \T \{ \phi_1(p_1) \phi_2(p_2) \} \rrangle.
\label{eq:partialtimeordering}
\end{equation}
The time-ordering operator is defined in position space by
\begin{equation}
	\T\{ \phi_1(x_1) \phi_2(x_2) \}
	= \Theta(x_1^0 - x_2^0) \phi_1(x_1) \phi_2(x_2)
	+ \Theta(x_2^0 - x_1^0) \phi_2(x_2) \phi_1(x_1).
\label{eq:timeordering:definition}
\end{equation}
Our notation for the operators differs from section~\ref{sec:scalar} because of the different physical interpretation of this correlation function. If $\bra{\phi_f(p_f)}$ still defines a final state created by a single operator, there is no notion of an initial state created by a local operator in the correlator~\eqref{eq:partialtimeordering}.
As a consequence, the momenta $p_1$ and $p_2$ might be time-like as well as space-like. The only requirement is that they add up to $-p_f$ that is time-like and has positive energy, by the condition~\eqref{eq:nullstates} on the final state.
The correlation function~\eqref{eq:partialtimeordering} is also obviously symmetric under the exchange of the operator $\phi_1$ and $\phi_2$, which means that it must be represented by a function that is symmetric under the simultaneous exchange of the momenta $p_1 \leftrightarrow p_2$ and of the scaling dimensions $\Delta_1 \leftrightarrow \Delta_2$.
This suggests the ansatz
\begin{equation}
	\llangle \phi_f(p_f) \T\{ \phi_1(p_1) \phi_2(p_2) \} \rrangle
	= \Theta(-p_f) (-p_f^2)^{(\Delta_f + \Delta_1 + \Delta_2 - 2d)/2}
	F_{12}\left( \frac{p_1^2}{p_f^2}, \frac{p_2^2}{p_f^2} \right),
\label{eq:ansatz:F12}
\end{equation}
in which the function $F_{12}$ enjoys the aforementioned symmetry. Since the partially-time-ordered 3-point function obeys the same conformal Ward identities as the Wightman function, it is possible to express the function $F_{12}$ as a linear combination of Appell $F_4$ functions of the type of eq.~\eqref{eq:4solutions}.
In other words, the two correlators are different solutions to the same system of partial differential equations, but with a different boundary condition.

As before, this boundary condition is provided by an OPE limit. Since the basis of momentum eigenstates created by single local operator insertion is complete, it must be possible to write
\begin{equation}
	\T \{ \phi_1(p_1) \phi_2(p_2) \} \ket{0} = \sum_\O \lambda_{\O12} \,
	\widetilde{C}_{\O12}^{\T}(p_1, p_1 + p_2)
	\ket{ \O(p_1 + p_2) },
\label{eq:OPE:momentum:T}
\end{equation}
for some function $\widetilde{C}_{\O12}^{\T}(p, q)$ that differs from $\widetilde{C}_{\O12}(p, q)$ of eq.~\eqref{eq:OPE:momentum}.
This function admits the formal expansion
\begin{equation}
	\widetilde{C}^{\T}_{\O12}(p, q)
	= \left[ 1 - i \frac{\Delta_\O + \Delta_1 - \Delta_2}{2 \Delta_\O} \, 
	q^\mu \frac{\partial}{\partial p^\mu} + \ldots \right]
	\int d^dx \, e^{i \, p \cdot x} (x^2 + i \epsilon)^{(\Delta_\O - \Delta_1 - \Delta_2)/2},
\label{eq:OPE:Ctilde:T}
\end{equation}
where the norm $(x^2 + i \epsilon)$ is the time-ordered analog of the norm appearing in eq.~\eqref{eq:OPE:Ctilde}.%
\footnote{One can write a similar OPE for the anti-time-ordered product of operators in which the sign of the $i \epsilon$ prescription is opposite.}
It corresponds to the time-ordered 2-point function of a fictitious operator with scaling dimension $(\Delta_1 + \Delta_2 - \Delta_\O)/2$.
Therefore, the integral on the right-hand-side of eq.~\eqref{eq:OPE:Ctilde:T} can be written as a momentum-space 2-point function,
\begin{equation}
	\llangle \T \{ \phi(-p) \phi(p) \} \rrangle
	= -i \frac{\pi^{d} \Gamma\left( \frac{d}{2} - \Delta \right)}
	{2^{2 \Delta - d} \Gamma\left( \Delta \right)}
	(p^2 - i \epsilon)^{\Delta - d/2}
	\equiv F_\Delta(p).
\label{eq:Feynman2ptfct}
\end{equation}
This gives immediately the asymptotic limit $p_f \to 0$ of the correlation function~\eqref{eq:partialtimeordering}:
\begin{equation}
\begin{aligned}
	\llangle \phi_f(p_f) \T\{ \phi_1(p_1) \phi_2(p_2) \} \rrangle_{p_f \to 0}
	&= -i
	\lambda_{f0i}
	\frac{2^{2d- \Delta_i - \Delta_0 - \Delta_f + 1} \pi^{d+1}
	\Gamma\left( \frac{\Delta_f - \Delta_1 - \Delta_2 + d}{2} \right)}
	{\Gamma\left( \Delta_f \right) \Gamma\left( \Delta_f - \frac{d-2}{2} \right)
	\Gamma\left( \frac{\Delta_1 + \Delta_2 - \Delta_f}{2} \right)}
	\\
	& \quad \times
	(-p_f^2)^{\Delta_f - d/2}
	(p_1^2 - i \epsilon)^{(\Delta_1 + \Delta_2 - \Delta_f - d)/2}.
\end{aligned}
\label{eq:limit:pf:zero:T}
\end{equation}
The symmetry $\phi_1 \leftrightarrow \phi_2$ is obvious on the right-hand side since $p_1^2 = p_2^2$ in this limit.
Unlike the Wightman function, the partially time-ordered 3-point function do not admit any other Lorentzian OPE limit.

\begin{figure}
	\centering
	\footnotesize
	\begin{tabular}{c@{\hspace{1.5cm}}c}
		\includegraphics[width=0.4\linewidth]{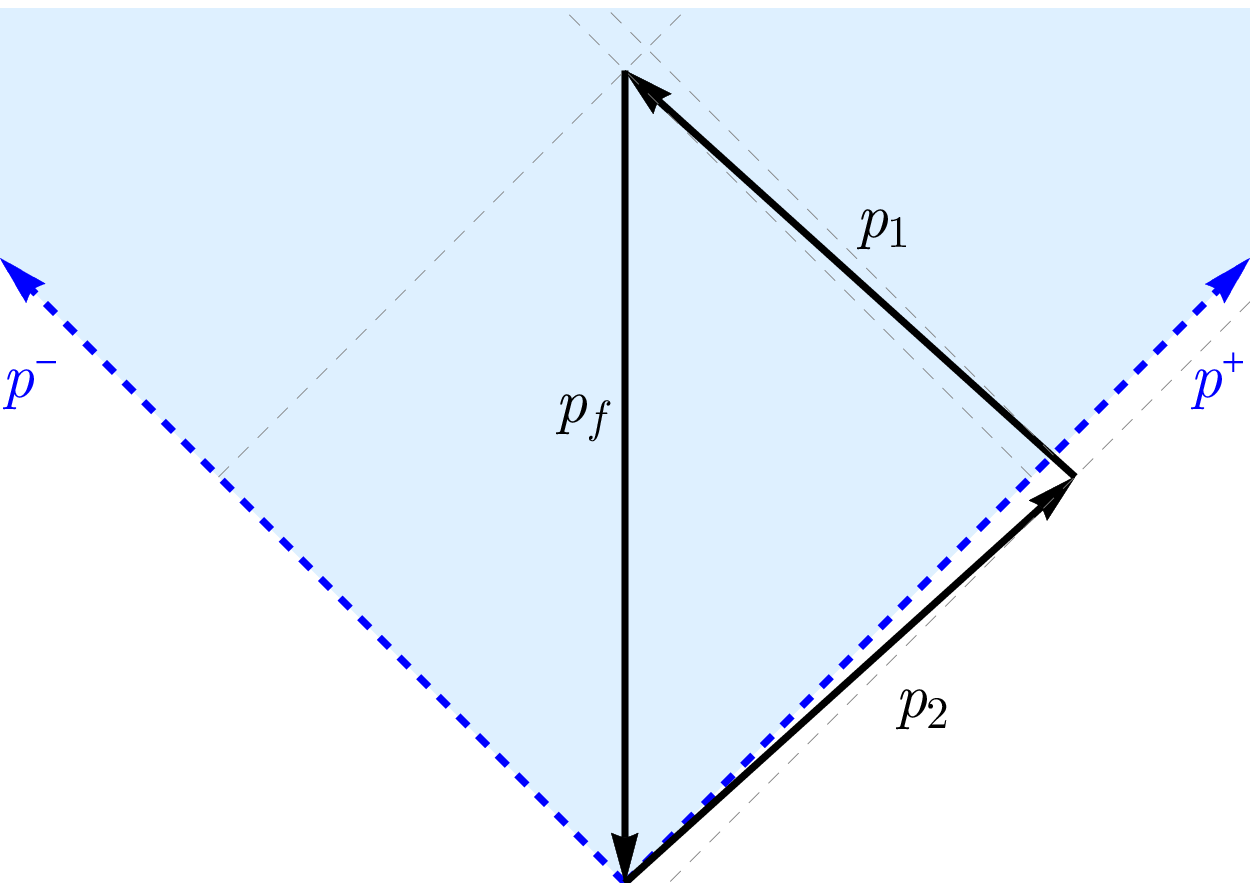}
		&
		\includegraphics[width=0.4\linewidth]{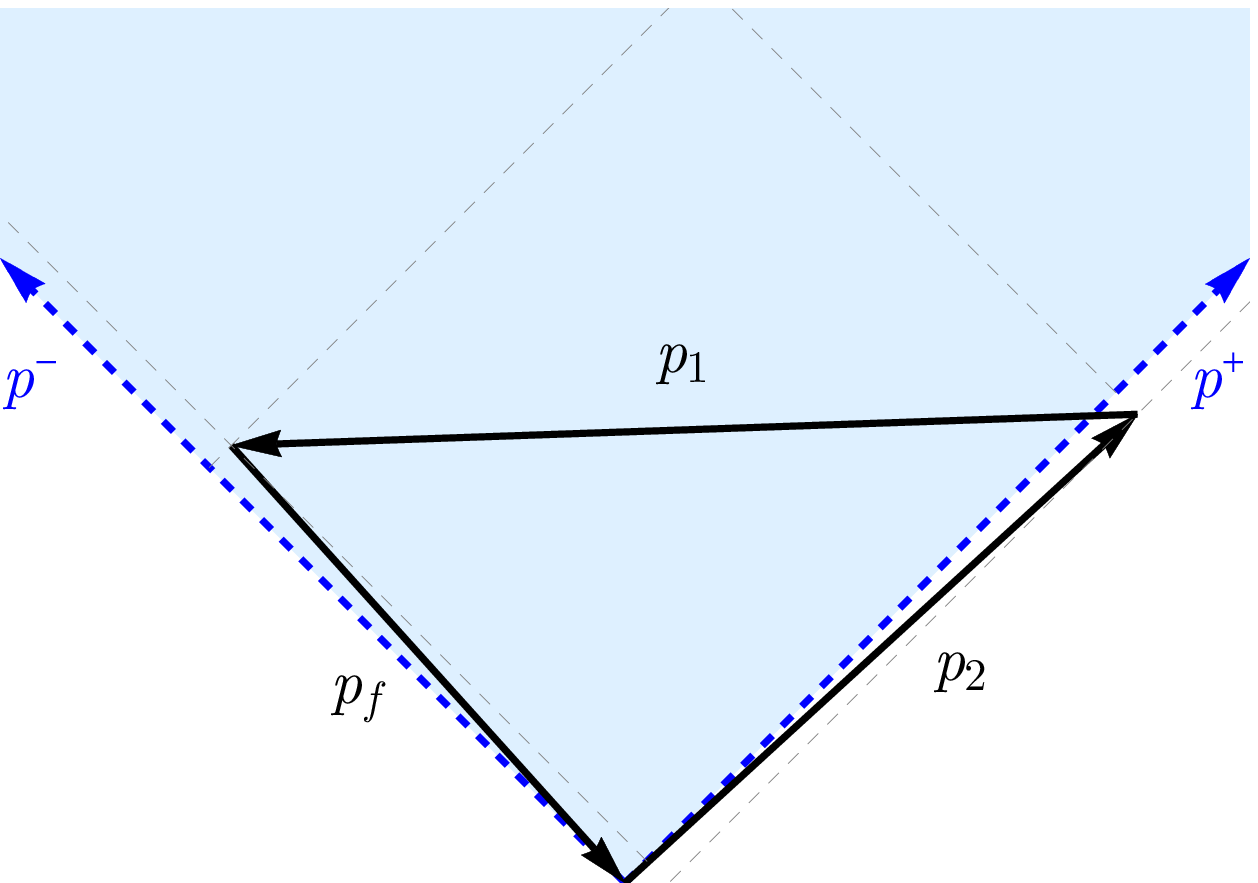}
		\\[1em]
		(a) $p_1^2, p_2^2 \to 0_+$ & (b) $p_1^2 \to 0_+$, $p_f^2 \to 0_-$
	\end{tabular}
	\caption{Examples of momentum configurations for the partially-time-ordered 3-point function~\eqref{eq:partialtimeordering}, which only has support when the momentum $-p_f$ lies in the light-cone shown in blue.
	In both examples the momenta $p_1$ and $p_2$ are space-like, and one can deform one configuration into the other without any light-cone crossing.
	}
	\label{fig:kinematics:partialtimeordering}
\end{figure}

It is not necessarily obvious how to reconcile this OPE limit with the symmetric ansatz~\eqref{eq:ansatz:F12} for the 3-point function. That ansatz gives a good description of the correlation function in a neighborhood of $p_1^2 = p_2^2 = 0$, which corresponds to a configuration of momenta as in figure~\ref{fig:kinematics:partialtimeordering}~(a). Both $p_1$ and $p_2$ lie close to the light-cone and have positive energy in that case, while the limit $p_f \to 0$ requires $p_2 \to -p_1$.
It is therefore useful to introduce the different ansatz
\begin{equation}
	\llangle \phi_f(p_f) \T\{ \phi_1(p_1) \phi_2(p_2) \} \rrangle
	= \Theta(-p_f) (p_1^2)^{(\Delta_f + \Delta_1 + \Delta_2 - 2d)/2}
	F_{f2}\left( \frac{p_f^2}{p_1^2}, \frac{p_2^2}{p_1^2} \right),
\label{eq:ansatz:Ff2}
\end{equation}
covering in particular configurations like figure~\ref{fig:kinematics:partialtimeordering}~(b) in which both $p_f^2$ and $p_2^2$ are small compared to $p_1^2$.
We will assume for now that both $p_1$ and $p_2$ space-like as in the figure and discuss later what happens when either one of them crosses a light-cone.
With this ansatz, the Ward identity~\eqref{eq:K:operator} for special conformal transformations implies that $F_{f2}(z_f, z_2)$ is a linear combination of the four functions
\begin{equation}
\begin{array}{c@{\qquad\quad}c}
	(-z_f)^{\Delta_f - d/2} (z_2)^{\Delta_2 - d/2} F_{\Delta_f \Delta_1 \Delta_2}(z_f, z_2),
	& (-z_f)^{\Delta_f - d/2} F_{\Delta_f \Delta_1 \Deltatilde_2}(z_f, z_2),
	\\
	(z_2)^{\Delta_2 - d/2} F_{\Deltatilde_f \Delta_1 \Delta_2}(z_f, z_2),
	& F_{\Deltatilde_f \Delta_1 \Deltatilde_2}(z_f, z_2),
\end{array}
\label{eq:4solutions:T}
\end{equation}
where $F_{\Delta_f \Delta_1 \Delta_2}$ is the Appell $F_4$ function given in eq.~\eqref{eq:F}.
Besides providing an explicit value for the limit $p_f \to 0$, the OPE \eqref{eq:OPE:Ctilde:T} also implies that the 3-point functions must scale like $(-p_f^2)^{\Delta_f - d/2}$ in the limit $p_f^2 \to 0_-$.
Among the 4 functions in eq.~\eqref{eq:4solutions:T}, only the first two follow this asymptotic behavior.
We must therefore have
\begin{equation}
	F_{f2}(z_f, z_2)
	=  (-z_f)^{\Delta_f - d/2} \left[ A \,(z_2)^{\Delta_2 - d/2} F_{\Delta_f \Delta_1 \Delta_2}(z_f, z_2)
	+ B \, F_{\Delta_f \Delta_1 \Deltatilde_2}(z_f, z_2) \right]
	\label{eq:F12}
\end{equation}
for some coefficients $A$ and $B$.
This form is readily compatible with the limit $p_f \to 0$, which corresponds to $z_f \to 0_-$ and $z_2 \to 1_-$.
The limit $z_f \to 0$ should be taken first, and one can then use eq.~\eqref{eq:F:limit:pf:zero} to obtain the limit $z_2 \to 1_-$.
For generic $A$ and $B$ there is a non-analytic piece proportional to $(1 - z_2)^{1 - \Delta_f}$ in this limit.
Requiring that this term vanishes and that the limit reproduces eq.~\eqref{eq:limit:pf:zero:T}, one obtains
\begin{equation}
\begin{aligned}
	A &= -i
	\lambda_{f0i}
	\frac{2^{2d- \Delta_i - \Delta_0 - \Delta_f + 1} \pi^{d+1}}
	{\Gamma\left( \Delta_f - \frac{d-2}{2} \right)
	\Gamma\left( \frac{\Delta_1 + \Delta_2 - \Delta_f}{2} \right)}
	\frac{\Gamma\left( \frac{d}{2} - \Delta_2 \right)}
	{\Gamma\left( \frac{\Delta_f + \Delta_1 - \Delta_2}{2} \right)},
	\\
	B &= -i
	\lambda_{f0i}
	\frac{2^{2d- \Delta_i - \Delta_0 - \Delta_f + 1} \pi^{d+1}}
	{\Gamma\left( \Delta_f - \frac{d-2}{2} \right)
	\Gamma\left( \frac{\Delta_1 + \Delta_2 - \Delta_f}{2} \right)}
	\frac{\Gamma\left( \Delta_2 - \frac{d}{2} \right)
	\Gamma\left( \frac{\Delta_f - \Delta_1 - \Delta_2 + d}{2} \right)}
	{\Gamma\left( \frac{\Delta_f - \Delta_1 + \Delta_2}{2} \right)
	\Gamma\left( \frac{\Delta_f + \Delta_1 + \Delta_2 - d}{2} \right)}.
\end{aligned}
\label{eq:AB}
\end{equation}
The symmetry $\phi_1 \leftrightarrow \phi_2$ is not at all obvious in this result. But one can now make use of the transformation property \eqref{eq:F4transformation} of the Appell $F_4$ function to bring the 3-point function in the form of the ansatz~\eqref{eq:ansatz:F12} where the symmetry becomes evident.
We find
\begin{equation}
\begin{aligned}
	F_{12}(z_1, z_2) &=-i
	\lambda_{f0i}
	\frac{2^{2d- \Delta_i - \Delta_0 - \Delta_f + 1} \pi^{d+1}}
	{\Gamma\left( \frac{\Delta_1 + \Delta_2 - \Delta_f}{2} \right)
	\Gamma\left( \frac{\Delta_f + \Delta_1 - \Delta_2}{2} \right)
	\Gamma\left( \frac{\Delta_f + \Delta_2 - \Delta_1}{2} \right)
	\Gamma\left( \frac{\Delta_f + \Delta_1 + \Delta_2 - d}{2} \right)}
	\\
	& \quad \times
	\bigg[
	f_{\Delta_f \Delta_1 \Delta_2} 
	(-z_1)^{\Delta_1 - d/2} (-z_2)^{\Delta_2 - d/2}
	F_{\Delta_1 \Delta_f \Delta_2}(z_1, z_2)
	\\
	& \quad \qquad
	+ f_{\Delta_f \Delta_1 \Deltatilde_2} 
	(-z_1)^{\Delta_1 - d/2} 
	F_{\Delta_1 \Delta_f \Deltatilde_2}(z_1, z_2)
	\\
	& \quad \qquad
	+ f_{\Delta_f \Deltatilde_1 \Delta_2} 
	(-z_2)^{\Delta_2 - d/2} 
	F_{\Deltatilde_1 \Delta_f \Delta_2}(z_1, z_2)
	\\
	& \quad \qquad
	+ f_{\Delta_f \Deltatilde_1 \Deltatilde_2} 
	F_{\Deltatilde_1 \Delta_f \Deltatilde_2}(z_1, z_2)
	\bigg],
\end{aligned}
\label{eq:partialtimeordering:spacelike}
\end{equation}
where we have denoted
\begin{equation}
	f_{\Delta_f \Delta_1 \Delta_2}
	= \frac{\Gamma\left( \frac{d}{2} - \Delta_1 \right)
	\Gamma\left( \frac{d}{2} - \Delta_2 \right)
	\Gamma\left( \frac{\Delta_f + \Delta_1 + \Delta_2 - d}{2} \right)}
	{\Gamma\left( 1 - \frac{\Delta_1 + \Delta_2 - \Delta_f}{2} \right)}.
\end{equation}
This is our result for the partially-time-ordered 3-point function.
The symmetry $\phi_1 \leftrightarrow \phi_2$ follows from the property $F_{\Delta_1 \Delta_f \Delta_2}(z_1, z_2) = F_{\Delta_2 \Delta_f \Delta_1}(z_2, z_1)$ of the Appell $F_4$ function.
The results of refs.~\cite{Gillioz:2016jnn, Gillioz:2018mto} are special cases of this expression. They correspond to the limit $p_1^2, p_2^2 \to 0_+$ which is finite under the assumption that $\Delta_1, \Delta_2 > \frac{d}{2}$.

Unlike the Wightman function, this correlation function is not analytic in the scaling dimensions. It has poles when $\Delta_1, \Delta_2 = \frac{d}{2} + n$ with $n \in \mathbb{N}$.
This is a well-known feature of the time-ordered correlation function, which is also present in the two-point function~\eqref{eq:Feynman2ptfct}: correlation functions involving operators with these special dimensions have anomalies and must be renormalized. In momentum space, this renormalization leads to the appearance of logarithms~\cite{Gillioz:2016jnn}.
The result \eqref{eq:partialtimeordering:spacelike} also shows that the 3-point function diverges when $\Delta_f = \Delta_1 + \Delta_2 - d - 2n$. We do not have an explanation for the presence of these poles.

\subsection{Relationship with the Wightman function}

The relationship between time-ordered and Wightman functions in momentum space is complicated.
There is no simple way to go from one to the other without invoking the position-space representation.
Nevertheless, there exists a link between the two given by the operator identity
\begin{equation}
	\T \{ \phi_1(p_1) \phi_2(p_2) \} + \Tbar \{ \phi_1(p_1) \phi_2(p_2) \}
	= \phi_1(p_1) \phi_2(p_2) + \phi_2(p_2) \phi_1(p_1),
\label{eq:timeorderingidentity}
\end{equation}
where $\Tbar$ indicates the reverse time-ordering operation.
When applied to the 2-point function, it implies that
\begin{equation}
	2 \re \llangle \T \{ \phi(-p) \phi(p) \} \rrangle
	= \llangle \phi(-p) \phi(p) \rrangle + \llangle \phi(p) \phi(-p) \rrangle,
\end{equation}
where have used the fact that the anti-time-ordered 2-point function is the complex conjugate of the time-ordered one.
This equality is satisfied by the 2-point functions given in eqs.~\eqref{eq:Wightman2ptfct} and \eqref{eq:Feynman2ptfct} for all values of the momentum $p$: when $p$ is space-like both Wightman functions vanish and the time-ordered function is purely imaginary; when $p$ is time-like, the non-trivial phase in the time-ordered function is precisely matched by the Wightman function.

The identity \eqref{eq:timeorderingidentity} become more interesting when we apply it to the 3-point function, as it provides an interesting verification of our result. We have
\begin{equation}
	2 \re \llangle \phi_f(p_f) \T\{ \phi_1(p_1) \phi_2(p_2) \} \rrangle
	= \llangle \phi_f(p_f) \phi_1(p_1) \phi_2(p_2) \rrangle
	+ \llangle \phi_f(p_f) \phi_2(p_2) \phi_1(p_1) \rrangle.
\label{eq:timeorderingidentity:3pt}
\end{equation}
In section~\ref{subsec:partialtimeordering} we focused on the regime where $p_1$ and $p_2$ are both space-like. In this case the Wightman functions on the right-hand side vanish, and we found indeed that the partially-time-ordered 3-point function \eqref{eq:partialtimeordering:spacelike} is purely imaginary.
When either $p_1$ or $p_2$ crosses the light-cone, necessarily with positive energy, then the corresponding Wightman function on the right-hand side becomes non-zero and equal to eq.~\eqref{eq:3scalars:spacelike}.
Thus the partially-time-ordered function must acquire a non-trivial phase.
The ansatz~\eqref{eq:ansatz:Ff2} is well-suited to study the case where $p_2$ crosses the light-cone.
Consistency with the light-cone limit \eqref{eq:limit:pf:zero:T} actually indicates that one should replace $p_2^2 \to p_2^2 - i \epsilon$ in that case, and one obtains therefore the relation
\begin{equation}
	2 \re \left[ e^{- i \pi (\Delta_2 - d/2)} A \right]
	= \lambdatilde_{f12}.
\end{equation}
It can be verified from the definitions \eqref{eq:AB} for $A$ and \eqref{eq:lambdatilde} for $\lambdatilde_{f12}$ that this is indeed satisfied.

Finally, it would be interesting to check the identity~\eqref{eq:timeorderingidentity:3pt} in the case where both momenta $p_1$ and $p_2$ are time-like. However, the analytic continuation of the partially-time-ordered function is ambiguous, and the information provided by the OPE is not sufficient to resolve it.
A naive guess would be to replace $z_1 \to z_1 + i \epsilon$ and $z_2 \to z_2 + i \epsilon$ in the expression~\eqref{eq:partialtimeordering:spacelike}. However, it can be verified that this guess does not satisfy~\eqref{eq:timeorderingidentity:3pt}: at least one additional term proportional to $z_1^{\Delta_1 - d/2} z_2^{\Delta_2 - d/2} F_{\Delta_1 \Delta_f \Delta_2}(z_1, z_2)$ must be present when both $p_1$ and $p_2$ are time-like.%
\footnote{Consider for instance that one can add to the function $F_{12}$ in eq.~\eqref{eq:F12} a term proportional to
\begin{equation*}
	\left[ (z_1 - i \epsilon)^{\Delta_1 - d/2} - (z_1 + i \epsilon)^{\Delta_1 - d/2} \right]
	\left[ (z_2 - i \epsilon)^{\Delta_2 - d/2} - (z_2 + i \epsilon)^{\Delta_2 - d/2} \right]
	F_{\Delta_1 \Delta_f \Delta_2}(z_1, z_2).
\end{equation*}
This term satisfies the conformal Ward identity, preserves the symmetry $\phi_1 \leftrightarrow \phi_2$ and vanishes whenever $p_1$ or $p_2$ is space-like. It might actually be the unique term with these properties.}
Note that the presence of this additional term is consistent with an observation that can be made using generalized free field theory: when the scaling dimensions satisfy $\Delta_f = \Delta_1 + \Delta_2$, the 3-point function is expected to factorize as
\begin{equation}
	\llangle{} [\phi_1\phi_2](p_f) \T\{ \phi_1(p_1) \phi_2(p_2) \} \rrangle
	= \lambda_{\phi_1\phi_2[\phi_1\phi_2]}
	\llangle \phi_1(-p_1) \phi_1(p_1) \rrangle
	\llangle \phi_2(-p_2) \phi_2(p_2) \rrangle.
\end{equation}
The right-hand side is non-zero when $p_1$ and $p_2$ are both time-like with positive energies.
However, the function $F_{12}$ in eq.~\eqref{eq:partialtimeordering:spacelike} vanishes identically when $\Delta_f = \Delta_1 + \Delta_2$.
Therefore it cannot be the complete answer when both $p_1$ and $p_2$ are time-like.
On the other hand, the other relation obtained from a generalized free field theory correlation function,
\begin{equation}
	\llangle \phi_f(p_f) \T\{ [\phi_f \phi_2](p_1) \phi_2(p_2) \} \rrangle
	= \lambda_{\phi_f\phi_2[\phi_f\phi_2]}
	\llangle \phi_f(p_f) \phi_f(-p_f) \rrangle
	\llangle \T\{ \phi_2(-p_2) \phi_2(p_2) \} \rrangle,
\end{equation}
is satisfied by eq.~\eqref{eq:F12} even when $p_2$ is time-like, provided that one makes the substitution $p_2^2 \to p_2^2 - i \epsilon$. In summary, the computation of the partially-time-ordered 3-point function in the regime of time-like $p_1$ and $p_2$ remains an interesting open problem that cannot be directly addressed with our method.

\subsection{The fully time-ordered 3-point function}

The last momentum-space 3-point function that one can consider is the fully time-ordered product
\begin{equation}
	\llangle \T \{ \phi_1(p_1) \phi_2(p_2) \phi_3(p_3) \} \rrangle.
\label{eq:fulltimeordering}
\end{equation}
It cannot be understood as the overlap of momentum eigenstates, and so the OPE analysis of the previous sections do not apply.
But the permutation symmetry $\phi_1 \leftrightarrow \phi_2 \leftrightarrow \phi_3$ between the 3 operators is actually sufficient to determine the 3-point function uniquely up to an overall coefficient:
when all three momenta are space-like, the Euclidean result of ref.~\cite{Coriano:2013jba} applies readily,%
\footnote{To fix the overall coefficient in this result, the authors of ref.~\cite{Coriano:2013jba} also use an OPE limit, but it is an Euclidean OPE that is conceptually different from the Lorentzian OPE discussed in section~\ref{subsec:OPElimits}.}
\begin{equation}
\begin{aligned}
	\llangle \T \{ \phi_1(p_1) \phi_2(p_2) \phi_3(p_3) \} \rrangle
	&= \frac{ \lambda_{123}  \, 2^{2d - \Delta_1 - \Delta_2 - \Delta_3} \pi^d}
	{\Gamma\left( \frac{\Delta_1 + \Delta_2 - \Delta_3}{2} \right)
	\Gamma\left( \frac{\Delta_1 + \Delta_3 - \Delta_2}{2} \right)
	\Gamma\left( \frac{\Delta_2 + \Delta_3 - \Delta_1}{2} \right)
	\Gamma\left( \frac{\Delta_1 + \Delta_2 + \Delta_3 - d}{2} \right)}
	\\
	& \quad \times
	(p_3^2)^{(\Delta_1 + \Delta_2 + \Delta_3 - 2d)/2}
	F_{\T}\left( \frac{p_1^2}{p_3^2}, \frac{p_2^2}{p_3^2} \right),
\label{eq:fulltimeordering:result}
\end{aligned}
\end{equation}
where
\begin{equation}
\begin{aligned}
	F_{\T}(z_1, z_2)
	&= 
	g_{\Delta_1 \Delta_3 \Delta_2}
	(z_1)^{\Delta_1 - d/2} (z_2)^{\Delta_2 - d/2}
	F_{\Delta_1 \Delta_3 \Delta_2}(z_1,z_2)
	\\
	& \quad
	+ g_{\Delta_1 \Delta_3 \Deltatilde_2}
	(z_1)^{\Delta_1 - d/2}
	F_{\Delta_1 \Delta_3 \Deltatilde_2}(z_1,z_2)
	\\
	& \quad
	+ g_{\Deltatilde_1 \Delta_3 \Delta_2}
	(z_2)^{\Delta_2 - d/2}
	F_{\Deltatilde_1 \Delta_3 \Delta_2}(z_1,z_2)
	\\
	& \quad
	+ g_{\Deltatilde_1 \Delta_3 \Deltatilde_2}
	F_{\Deltatilde_1 \Delta_3 \Deltatilde_2}(z_1,z_2),
\end{aligned}
\label{eq:FT}
\end{equation}
and the coefficients $g_{\Delta_1 \Delta_3 \Delta_2}$ are defined by
\begin{equation}
	g_{\Delta_1 \Delta_3 \Delta_2}
	\equiv \Gamma\left( \tfrac{d}{2} - \Delta_1 \right) \Gamma\left( \tfrac{d}{2} - \Delta_2 \right)
	\Gamma\left( \tfrac{\Delta_1 + \Delta_2 - \Delta_3}{2} \right)
	\Gamma\left( \tfrac{\Delta_1 + \Delta_2 - \Deltatilde_3}{2} \right).
\end{equation}
Once again the 3-point function is expressed in terms of the solutions to the conformal Ward identities~\eqref{eq:K:operator}. This time all four solutions appear in the 3-point function, no matter which of the momenta is taken as the reference momentum.

This fully-time-ordered 3-point function can be related to the partially-time-ordered correlator using an identity similar to eq.~\eqref{eq:timeorderingidentity}:
\begin{equation}
\begin{aligned}
	\T\{ \phi_1 \phi_2 \phi_3 \} - \Tbar\{ \phi_1 \phi_2 \phi_3 \}
	&= \phi_1 \T\{ \phi_2 \phi_3 \} 
	+ \phi_2 \T\{ \phi_1 \phi_3 \} 
	+ \phi_3 \T\{ \phi_1 \phi_2 \} 
	\\
	& \quad\quad
	- \Tbar\{ \phi_1 \phi_2 \} \phi_3
	- \Tbar\{ \phi_1 \phi_3 \} \phi_2
	- \Tbar\{ \phi_2 \phi_3 \} \phi_1.
\end{aligned}
\label{eq:fulltimeorderingidentity}
\end{equation}
We have omitted to write the argument of the operators because this identity is purely combinatoric and applies in position space as well as in momentum space.
In position space it can be verified by expanding the time-ordered product according to its definition.
In momentum space, the content of the identity is more interesting as it provides an interesting relation between eq.~\eqref{eq:fulltimeordering:result} and the results of the previous sections.
When all three momenta are space-like, each of the 3-point functions on the right-hand side vanishes individually as it involves a null state. To see that the left-hand side vanishes as well, it is sufficient to note that the time-ordered and anti-time-ordered products are related by complex conjugation,%
\footnote{This can be seen in position space, where the time-ordered and anti-time-ordered correlation functions only differ by the sign of the Feynman $i \epsilon$ prescription. Since both functions are symmetric under $x \to -x$, the Fourier transform preserves this property.}
so that the identity becomes
\begin{equation}
	\im \llangle \T\{ \phi_1(p_1) \phi_2(p_2) \phi_3(p_3) \} \rrangle
	= 0
	\qquad
	(p_1^2, p_2^2, p_3^2 > 0).
\end{equation}
The function $F_T$ in eq.~\eqref{eq:FT} is indeed real. This means that there is no distinction between the time-ordered and anti-time-ordered products when all momenta are space-like.

If one of the momenta is instead time-like (say $p_1$), one term on the right-hand side of the identity~\eqref{eq:fulltimeorderingidentity} is non-zero. In the time-ordered 3-point function, this configuration can be reached by analytic continuation in $p_1^2$. There are only two possible analytic continuations of the function $F_T$ that preserve the permutation symmetry of the operators: one of them consist in taking $p_1^2 \to p_1^2 - i \epsilon$ and the other 
$p_1^2 \to p_1^2 + i \epsilon$. From the representation of the 3-point function as the Fourier transform of the position-space correlator, it is easy to see that the first of these analytic continuations corresponds to the time-ordered product and the second to the anti-time-ordered product.
They are the complex conjugate of each other, and thus one can write
\begin{equation}
\begin{aligned}
	2 i \im \llangle \T\{ \phi_1(p_1) \phi_2(p_2) \phi_3(p_3) \} \rrangle
	&= \llangle \phi_1(p_1) \T\{ \phi_2(p_2) \phi_3(p_3) \} \rrangle
	\\
	& \quad
	+  \llangle \Tbar\{ \phi_2(p_2) \phi_3(p_3) \} \phi_1(p_1) \rrangle
	\qquad
	(p_2^2, p_3^2 > 0).
\end{aligned}
\end{equation}
When $p_1$ has positive energy, the right-hand side is given by eq.~\eqref{eq:ansatz:F12}, or equivalently eq.~\eqref{eq:ansatz:Ff2}.
It can be verified that these expressions precisely match the imaginary part of eq.~\eqref{eq:fulltimeordering:result} with $p_1^2 \to p_1^2 - i \epsilon$.
This provides a simple and yet non-trivial verification of the results of section~\ref{subsec:partialtimeordering}.


\section{Discussion}

In this paper, we have provided a simple closed-form%
\footnote{The terminology of ``closed-form'' is used because the Appell $F_4$ function is a recognized function whose properties are known and well-documented, but really it does not carry much meaning:
to evaluate it, one must either use one of its integral representations or the double-infinite hypergeometric series~\eqref{eq:AppellF4}.
Nevertheless, we have provided for every possible momentum configurations at least one representation in terms of which the double hypergeometric series converges, and thus the function can be approximated to arbitrary precision with a truncated series.}
expression for the momentum-space Wightman function of 3 scalar operators, as well as two scalars and one traceless symmetric tensor with arbitrary spin.
Besides the explicit results, we have given a detailed explanation of the logic underlying this approach so that the interested readers might themselves proceed to the derivation of correlation functions not given here.
This method should be particularly suited to study the Wightman correlation functions of operators such as conserved currents or the energy-momentum tensor. For other spin representations, one might want to develop the theory of weight-shifting operators in momentum space.

In addition, studying the (partially-)time-ordered correlation function in the case of scalar operators, we have found that they take generically a more complicated form, which we were not able to fix completely by analytic continuation away from the OPE limits.
One observes also that the time-ordered correlation functions can be expressed as a sum of the Wightman function and of some its shadow transforms.
This supports the idea that the Wightman function are really the building blocks in a Lorentzian conformal field theory.

Finally, we would like to emphasize that the computation of the Wightman 3-point function opens the door to the computation of higher-point functions through the \linebreak momentum-space OPE:
one way of interpreting our results is in terms of the relation
\begin{equation}
	\phi_1(p_1) \phi_2(p_2) \ket{0}
	= \sum_\O \lambda_{\O12} \widetilde{C}_{\O12}(p_1, p_1 + p_2) \O(p_1 + p_2) \ket{0},
\end{equation}
which we have formally established for any operator $\O$ and in every kinematic configuration of the momenta $p_1$ and $p_2$.%
\footnote{In practice determining $\widetilde{C}_{\O12}(p_1, p_1 + p_2)$ from the 3-point function still requires multiplication with the inverse of the 2-point function of the operator $\O$, which for operators of large spin can be a bit tedious.}
Many questions about the convergence of this OPE in a 4-point function remain to be answered:
it is guaranteed to converge in a distributional sense only~\cite{Mack:1976pa}, but not necessarily at every single point in momentum space. The problem has been addressed in $d = 2$~\cite{Gillioz:2019iye}, but it remains open in higher dimensions.

\subsection*{Acknowledgments}

The author would like to thank
Brian Henning,
Xiaochuan Lu,
Markus Luty,
Guram Mikaberidze,
Marco Meineri,
and Jo\~ao Penedones
for discussions.
The work of the author at EPFL was supported by the Swiss National Science Foundation through the NCCR SwissMAP.


\appendix
\section{Conformal algebra}
\label{app:algebra}

In this appendix we describe our conventions for the generators of the Lorentzian conformal group $\SO(d,2)$ and the infinitesimal transformation of the operators in the momentum-space representation.
There are $d (d+1) / 2$ generators of $\SO(d,2)$, denoted by the antisymmetric tensors $J^{AB}$ with indices $A,B =  0, \ldots, d+1$. They are hermitian, $(J^{AB})^\dag = J^{AB}$, and obey the algebra
\begin{equation}
	[ J^{AB}, J^{CD} ]
	= -i \left( \eta^{AC} J^{BD} - \eta^{AD} J^{BC}
	- \eta^{BC} J^{AD} + \eta^{BD} J^{AC} \right),
\end{equation}
with metric $\eta^{AB} = \text{diag}( -1, +1, \ldots, +1, -1)$.
We take the Lorentz indices in the range $0, \ldots, d-1$ and identify
\begin{equation}
	M^{\mu\nu} = J^{\mu\nu},
	\qquad
	P^\mu = J^{\mu \, d} - J^{\mu \, d+1},
	\qquad
	K^\mu = J^{\mu \, d} + J^{\mu \, d+1},
	\qquad
	D = J^{d \, d+1}.
\end{equation}
This gives the commutation relations%
\footnote{Our conventions match refs.~\cite{Simmons-Duffin:2016gjk, Poland:2018epd} with an additional $i$ to have Hermitian generators and hence unitary representations in Lorentzian signature.
}
\begin{equation}
\begin{aligned}[]
	[ M^{\mu\nu}, M^{\rho\sigma} ]
	&= -i \left( \eta^{\mu\rho} M^{\nu\sigma} - \eta^{\mu\sigma} M^{\nu\rho}
	- \eta^{\nu\rho} M^{\mu\sigma} + \eta^{\nu\sigma} M^{\mu\rho} \right),
	\\
	[ M^{\mu\nu}, P^\rho ] &= -i \left( \eta^{\mu\rho} P^\nu - \eta^{\nu\rho} P^\mu \right),
	\\
	[ M^{\mu\nu}, K^\rho ] &= -i \left( \eta^{\mu\rho} K^\nu - \eta^{\nu\rho} K^\mu \right),
	\\
	[ D, P^\mu ] &= i P^\mu,
	\\
	[ D, K^\mu ] &= -i K^\mu,
	\\
	[ P^\mu, K^\nu ] &= -2 i \left( \eta^{\mu\nu} D + M^{\mu\nu} \right),
\end{aligned}
\label{eq:conformalalgebra}
\end{equation}
and all other commutators vanish.
The transformation rules for a primary operator $\O(p)$ with scaling dimension $\Delta$ are
\begin{equation}
\begin{aligned}
	\left[ P^\mu, \O(p) \right] &= p^\mu \O(p),
	\\
	\left[ D, \O(p) \right] &= i \left( p^\rho \frac{\partial}{\partial p^\rho} +  d - \Delta \right) \O(p),
	\\
	\left[ M^{\mu\nu}, \O(p) \right] &= i \left( p^\mu \frac{\partial}{\partial p^\nu}
	- p^\nu \frac{\partial}{\partial p^\mu} - \Sigma^{\mu\nu} \right)  \O(p),
	\\
	\left[ K^\mu, \O(p) \right] &= \left( -2 p^\rho \frac{\partial^2}{\partial p^\mu \partial p^\rho}
	+ p^\mu \frac{\partial^2}{\partial p^\rho \partial p^\rho}
	+ 2 (\Delta - d) \frac{\partial}{\partial p^\mu}
	- 2\frac{\partial}{\partial p^\rho} \Sigma^{\mu\rho} \right) \O(p),
\end{aligned}
\label{eq:primaryfieldtransformations}
\end{equation}
where $\Sigma^{\mu\nu}$ is the spin matrix acting on the indices of the operators $\O$ which are implicit here.
These transformations follow from the definition~\eqref{eq:operator:momentum} of operators in momentum space, together with the decomposition of the Hilbert space into irreducible representations of the Lorentz group and dilatations,
\begin{equation}
	D \ket{ \O} = - i \Delta \ket{ \O },
	\qquad
	M^{\mu\nu} \ket{ \O} = - i \Sigma^{\mu\nu} \ket{ \O },
	\qquad
	K^\mu \ket{ \O} = 0.
\end{equation}


\section{Direct Fourier transform}
\label{app:Fourierintegrals}

Consider the position-space Wightman 3-point function of scalar operators
\begin{equation}
	\bra{0} \O_f(x_f) \O_0(x_0) \O_i(x_i) \ket{0}
	= \frac{\lambda_{f0i}}
	{(x_{f0}^2)^{(\Delta_f + \Delta_0 - \Delta_i)/2}
	(x_{fi}^2)^{(\Delta_f + \Delta_i - \Delta_0)/2}
	(x_{0i}^2)^{(\Delta_0 + \Delta_i - \Delta_f)/2}},
\label{eq:positionspace3ptfct}
\end{equation}
where we have denoted $x_{ab}^2 = -(x_a^0 - x_b^0 - i \epsilon)^2 + (\vec{x}_a - \vec{x}_b)^2$.
The goal of this appendix is to bring the Fourier transform of this expression in a form where it can easily be evaluated numerically, and to derive results analytically in limits where the integrals are tractable.

Since eq.~\eqref{eq:positionspace3ptfct} is the product of three Wightman function, its Fourier transform can be written
\begin{equation}
	\llangle \O_f(p_f) \O_0(p_0) \O_i(p_i) \rrangle
	= \lambda_{f0i} \int \frac{d^dk}{(2\pi)^d}
	W_\alpha(k)
	W_\beta(-p_f - k)
	W_\gamma(p_i - k)
\label{eq:triangle}
\end{equation}
where 
\begin{equation}
	W_\alpha(k) = \int d^dx \, \frac{e^{i \, k \cdot x}}
	{\left[ -(x^0 - i \epsilon)^2 + \vec{x}^2 \right]^\alpha}
\end{equation}
and we have defined
\begin{equation}
	\alpha = \frac{\Delta_f + \Delta_i - \Delta_0}{2},
	\qquad
	\beta = \frac{\Delta_f + \Delta_0 - \Delta_i}{2},
	\qquad
	\gamma = \frac{\Delta_i + \Delta_0 - \Delta_f}{2}.
\end{equation}
We will denote the integral in \eqref{eq:triangle} with $W_{\alpha\beta\gamma}(p_f, p_i)$.
Using the formula \eqref{eq:Wightman2ptfct} for $W_\alpha(k)$,
\begin{equation}
\begin{aligned}
	W_{\alpha\beta\gamma}(p_f, p_i) &= 
	\frac{2^{2d - 2 \alpha - 2 \beta - 2 \gamma + 3} \pi^{(d+6)/2}}
	{\Gamma(\alpha) \Gamma(\beta) \Gamma(\gamma)
	\Gamma\left( \alpha - \frac{d-2}{2} \right)
	\Gamma\left( \beta - \frac{d-2}{2} \right)
	\Gamma\left( \gamma - \frac{d-2}{2} \right)}	
	\\
	& \quad \times
	\int d^dk \,
	\Theta(k) \Theta(-p_f - k) \Theta(p_i - k)
	\\
	& \quad \hspace{17mm} \times
	\left[ -k^2 \right]^{\alpha - d/2}
	\left[ -(-p_f - k)^2 \right]^{\beta - d/2}
	\left[ -(p_i - k)^2 \right]^{\gamma - d/2}
\end{aligned}
\label{eq:W}
\end{equation}
This integral is free of ultraviolet divergences since the region of integration in $k$ is bounded by the presence of the $\Theta$ functions defined in eq.~\eqref{eq:Theta}.
It can however have infrared divergences depending on the value of the parameters $\alpha$, $\beta$ and $\gamma$ and on the kinematics.
To avoid this situation we will assume that the momenta $p_i$ and $p_f$ are non-colinear and that
\begin{equation}
	\alpha, \beta, \gamma > \frac{d}{2}.
\label{eq:convergenceassumption}
\end{equation}
Working in $d > 2$ space-time dimensions, it is convenient to introduce light-cone coordinates $k = (k^+, k^-, \vec{k}^\perp)$, such that the scalar product of two vectors is $k_1 \cdot k_2 = - \frac{1}{2} \left( k_1^+ k_2^- + k_1^- k_2^+ \right) + \vec{k}_1^\perp \cdot \vec{k}_2^\perp$ and the integration measure $d^dk = \frac{1}{2} dk^+ dk^- d^{d-2}\vec{k}^\perp$.
This corresponds to choosing $k^0 = \frac{1}{2} \left( k^+ + k^- \right)$ and $k^{\|} = \frac{1}{2} \left( k^+ - k^- \right)$, where $k^{\|}$ indicates some preferred space direction.
We can take in particular this direction in the plane of $p_f$ and $p_i$, so that $p_f = (p_f^+, p_f^-, 0)$ and $p_i = (p_i^+, p_i^-, 0)$.
Now the $\Theta$ functions give the conditions
\begin{equation}
	0 < k^+ < \min( -p_f^+, p_i^+) \equiv k_\text{max}^+,
	\qquad
	0 < k^- < \min( -p_f^-, p_i^-) \equiv k_\text{max}^-,
\label{eq:kplusminusmax}
\end{equation}
and
\begin{equation}
	|k^\perp| < \min\left( \sqrt{ k^+ k^-},
	\sqrt{(-p_f^+ - k^+) (-p_f^- - k^-)}, 
	\sqrt{(p_i^+ - k^+) (p_i^- - k^-)} \right)
	\equiv k_\text{max}^\perp.
\label{eq:kperpmax}
\end{equation}
It is immediately obvious that the integral is non-zero if $p_f^\pm < 0$ and $p_i^\pm > 0$ only, i.e.~the result will be proportional to $\Theta(-p_f) \Theta(p_i)$ as expected.
Using spherical coordinates for $k^\perp$, we have
\begin{equation}
	W_{\alpha\beta\gamma}(p_f, p_i)
	= \frac{2^{2d - 2 \alpha - 2 \beta - 2 \gamma + 2} \pi^{d+2} \,
	\widetilde{W}_{\alpha\beta\gamma}(p_f, p_i)}
	{\Gamma\left( \frac{d-2}{2} \right)
	\Gamma(\alpha) \Gamma(\beta) \Gamma(\gamma)
	\Gamma\left( \alpha - \frac{d-2}{2} \right)
	\Gamma\left( \beta - \frac{d-2}{2} \right)
	\Gamma\left( \gamma - \frac{d-2}{2} \right)}
\end{equation}
where we have now defined
\begin{equation}
\begin{aligned}
	\widetilde{W}_{\alpha\beta\gamma}(p_f, p_i)
	= 2
	\int_0^{k_\text{max}^+} dk^+
	\int_0^{k_\text{max}^-} dk^-
	\int_0^{k_\text{max}^\perp} dk^\perp \,
	(k^\perp)^{d-3}
	\left[ k^+ k^- - (k^\perp)^2 \right]^{\alpha - d/2}
	\\
	\times \left[ (-p_f^+ - k^+) (-p_f^- - k^-) - (k^\perp)^2 \right]^{\beta - d/2} &
	\\
	\times \left[ (p_i^+ - k^+) (p_i^- - k^-) - (k^\perp)^2 \right]^{\gamma - d/2} &.
\end{aligned}
\label{eq:Wtilde}
\end{equation}
In this form, the integral is easy to evaluate numerically, but still hard to handle analytically. 
Besides numerical checks of the results of section~\ref{sec:scalar} that have been performed, we consider two  kinematic limits in which it can be evaluated explicitly.

The first limit is
\begin{equation}
	-p_f^-, p_i^+ \ll -p_f^+, p_i^-.
\end{equation}
In this case $p_0 = -p_f - p_i$ is space-like since $p_0^2 \approx - p_f^+ p_i^- > 0$.
We are therefore in the situation of eq.~\eqref{eq:limit:massless},
\begin{equation}
	-p_i^2, -p_f^2 \ll p_0^2.
\end{equation}
The integral \eqref{eq:Wtilde} simplifies to
\begin{equation}
\begin{aligned}
	\widetilde{W}_{\alpha\beta\gamma}(p_f, p_i)
	\approx 2
	\int_0^{p_i^+} dk^+
	\int_0^{-p_f^-} dk^-
	\int_0^{\sqrt{k^+ k^-}} dk^\perp \,
	(k^\perp)^{d-3}
	\left[ k^+ k^- - (k^\perp)^2 \right]^{\alpha - d/2}
	\\
	\times \left[ -p_f^+ (-p_f^- - k^-) \right]^{\beta - d/2}
	\left[ (p_i^+ - k^+) p_i^- \right]^{\gamma - d/2} &.
\end{aligned}
\end{equation}
With the change of variable $k^+ = p_i^+ u$, $k^- = -p_f^- v$, $k^\perp = (-p_f^- p_i^+ w)^{1/2}$, the dependence on the momenta can be factored out,
\begin{equation}
\begin{aligned}
	\widetilde{W}_{\alpha\beta\gamma}(p_f, p_i)
	&\approx
	(-p_f^- p_i^+)^\alpha
	(p_f^+  p_f^-)^{\beta - d/2}
	(p_i^+ p_i^-)^{\gamma - d/2}
	\\
	& \quad \times 
	\int_0^1 du
	\int_0^1 dv
	\int_0^{u v} dw \,
	w^{(d-4)/2} (u v - w)^{\alpha - d/2}
	(1-v)^{\beta - d/2}
	(1-u)^{\gamma - d/2}.
\end{aligned}
\end{equation}
After rescaling $w \to u v w$, the three integrals factorize and one arrives at
\begin{equation}
\begin{aligned}
	\widetilde{W}_{\alpha\beta\gamma}(p_f, p_i)
	&\approx
	(-p_f^- p_i^+)^\alpha
	(p_f^+  p_f^-)^{\beta - d/2}
	(p_i^+ p_i^-)^{\gamma - d/2}
	\\
	& \quad \times
	\frac{\Gamma\left( \frac{d-2}{2} \right)
	\Gamma(\alpha)
	\Gamma\left(\alpha - \frac{d-2}{2} \right)
	\Gamma\left(\beta - \frac{d-2}{2} \right)
	\Gamma\left(\gamma - \frac{d-2}{2} \right)}
	{\Gamma\left( \alpha + \beta - \frac{d-2}{2} \right)
	\Gamma\left( \alpha + \gamma - \frac{d-2}{2} \right)},
\end{aligned}
\end{equation}
or for the complete integral \eqref{eq:W},
\begin{equation}
	W_{\alpha\beta\gamma}(p_f, p_i)
	\approx \frac{2^{2d - 2 \alpha - 2 \beta - 2 \gamma + 2} \pi^{d+2}}
	{\Gamma(\beta) \Gamma(\gamma)
	\Gamma\left( \alpha + \beta - \frac{d-2}{2} \right)
	\Gamma\left( \alpha + \gamma - \frac{d-2}{2} \right)}
	\frac{(-p_f^2)^{\alpha + \beta - d/2}
	(-p_i^2)^{\alpha + \gamma - d/2}}
	{(p_0^2)^\alpha}.
\end{equation}
This is in agreement with eq.~\eqref{eq:3scalars:spacelike} and the definition \eqref{eq:lambdatilde} of $\lambdatilde_{f0i}$.
Note that the singularity in $d = 2$ that appears at intermediate steps of the computation is absent in the final limit; one can verify that a derivation using light-cone coordinates in $d = 2$ (i.e.~without the orthogonal component $k^\perp$) gives an identical result.
Similarly, this limit is completely analytic in $\alpha$, $\beta$ and $\gamma$ so that the assumption \eqref{eq:convergenceassumption} can be relaxed.

The second limit that can be taken analytically is
\begin{equation}
	-p_f^\pm \ll p_i^\pm.
\end{equation}
It corresponds to
\begin{equation}
	-p_f^2 \ll -p_i^2 \approx -p_0^2
	\qquad
	(p_0^2 < 0).
\end{equation}
In this case the integral \eqref{eq:Wtilde} can be approximated with
\begin{equation}
\begin{aligned}
	\widetilde{W}_{\alpha\beta\gamma}(p_f, p_i)
	\approx 2
	\int_0^{-p_f^+} dk^+
	\int_0^{-p_f^-} dk^-
	\int_0^{k_\text{max}^\perp} dk^\perp \,
	(k^\perp)^{d-3}
	\left[ k^+ k^- - (k^\perp)^2 \right]^{\alpha - d/2}
	\\
	\times \left[ (-p_f^+ - k^+) (-p_f^- - k^-) - (k^\perp)^2 \right]^{\beta - d/2} 
	\left[ p_i^+ p_i^- \right]^{\gamma - d/2} &,
\end{aligned}
\end{equation}
which after the change of variables $k^+ = -p_f^+ u$, $k^- = -p_f^- v$, $k^\perp = (p_f^+ p_f^- w)^{1/2}$ becomes
\begin{equation}
\begin{aligned}
	\widetilde{W}_{\alpha\beta\gamma}(p_f, p_i)
	&\approx
	(p_f^+ p_f^-)^{\alpha + \beta - d/2}
	(p_i^+ p_i^-)^{\gamma - d/2}
	\int_0^1 du
	\int_0^1 dv
	\int_0^{\min[ uv, (1-u)(1-v)]} dw 
	\\
	& \quad \hspace{30mm}
	\times w^{(d-4)/2}
	\left[ u v - w \right]^{\alpha - d/2}
	\left[ (1 - u) (1 - v) - w \right]^{\beta - d/2}.
\end{aligned}
\end{equation}
To evaluate the remaining integral, one performs the change of variable
\begin{equation}
	u = 1 - \chi \xi,
	\qquad
	v = \eta \xi,
	\qquad
	w = \chi \eta \xi (1 - \xi),
\label{eq:changeofvariable}
\end{equation}
in terms of which
\begin{equation}
	\int_0^1 du
	\int_0^1 dv
	\int_0^{\min[uv, (1-u)(1-v)]} dw
	= \int_0^1 d\chi \, \chi
	\int_0^1 d\eta \, \eta
	\int_0^1 d\xi \, \xi^2.
\end{equation}
This gives again three independent integrals that can be expressed as ratios of $\Gamma$-functions, and we find
\begin{equation}
	\widetilde{W}_{\alpha\beta\gamma}(p_f, p_i)
	\approx
	(p_f^+ p_f^-)^{\alpha + \beta - d/2}
	(p_i^+ p_i^-)^{\gamma - d/2}
	\frac{\Gamma\left( \frac{d-2}{2} \right)
	\Gamma(\alpha) \Gamma(\beta)
	\Gamma\left(\alpha - \frac{d-2}{2} \right)
	\Gamma\left(\beta - \frac{d-2}{2} \right)}
	{\Gamma(\alpha + \beta)
	\Gamma\left( \alpha + \beta - \frac{d-2}{2} \right)}.
\end{equation}
This gives finally for the integral \eqref{eq:W}
\begin{equation}
	W_{\alpha\beta\gamma}(p_f, p_i)
	\approx \frac{2^{2d - 2 \alpha - 2 \beta - 2 \gamma + 2} \pi^{d+2}}
	{\Gamma(\alpha + \beta) \Gamma(\gamma)
	\Gamma\left( \alpha + \beta - \frac{d-2}{2} \right)
	\Gamma\left( \gamma - \frac{d-2}{2} \right)}
	(-p_i^2)^{\gamma - d/2}
	(-p_f^2)^{\alpha + \beta - d/2}.
\end{equation}
Again, we find perfect agreement with the OPE limit \eqref{eq:limit:pf:zero}.

\bibliography{Bibliography}
\bibliographystyle{utphys}

\end{document}